\pdfoutput=1

\documentclass{article}

\PassOptionsToPackage{numbers}{natbib}
\usepackage[final]{neurips_data_2024}

\usepackage[T1]{fontenc}
\usepackage[hidelinks]{hyperref}
\usepackage{url}
\usepackage{booktabs}
\usepackage{amsfonts}
\usepackage{nicefrac}
\usepackage[svgnames,table]{xcolor}

\usepackage{enumitem}
\usepackage[most]{tcolorbox}
\usepackage{graphicx}
\usepackage{wrapfig}
\usepackage{subcaption}
\usepackage{placeins}
\usepackage{wasysym}
\usepackage{adjustbox}
\usepackage{cleveref}
\usepackage{multirow}
\usepackage{minitoc}
\usepackage{ifthen}
\usepackage[textsize=tiny]{todonotes}
\usepackage{etoolbox} 
\usepackage{pgf}
\usepackage{pgfplots}
\usepackage{colortbl}
\pgfplotsset{compat=1.16}
\usepackage{pgffor}
\usepackage{array}
\usepackage{rotating}
\usepackage{svg}
\usepackage{amsmath}
\usepackage{twemojis}
\usepackage{tabularx}
\usepackage{array}
\usepackage{fancyhdr}
\usepackage{fontawesome5}

\usepackage[final]{microtype}
\DeclareRobustCommand{\suppinfo}[1]{%
    \ifthenelse{\boolean{includeappendix}}%
    {\Cref{#1}}%
    {Supplementary Information (SI)}%
}
\newboolean{includeappendix}
\setboolean{includeappendix}{true}

\usepackage{lscape}
\usepackage{pifont}
\newcommand\tldrDone[1]{}

\newlength{\myMheight}
\newcommand{\hf}{\includegraphics[height=\myMheight]{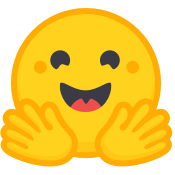}}

\newcommand\headsample{\textbf{Head Sample}}

\newcommand\modelpopulation{\textbf{Population Sample}}
\newcommand\rollingwindowfilterbf{\textbf{Rolling Window Filter}}
\newcommand\rollingwindowfilter{Rolling Window Filter}
\newcommand\derivedAuthorbf{\textbf{Recursive Model Attribution}}
\newcommand\derivedAuthor{Recursive Model Attribution}

\definecolor{greencell}{HTML}{2e7d43}
\definecolor{redcell}{HTML}{FF3333}

\definecolor{strongred}{RGB}{215, 48, 39}      
\definecolor{softred}{RGB}{244, 109, 67}       
\definecolor{lightred}{RGB}{253, 174, 107}     
\definecolor{verylightred}{RGB}{254, 224, 210} 
\definecolor{neutral}{RGB}{247, 247, 247}      
\definecolor{verylightgreen}{RGB}{230, 245, 208} 
\definecolor{lightgreen}{RGB}{161, 217, 155}   
\definecolor{softgreen}{RGB}{82, 190, 128}     
\definecolor{stronggreen}{RGB}{27, 120, 55}    

\newcommand{\gradientcell}[2]{%
  \pgfmathsetmacro{\val}{#1}
  \ifdim\val pt<-0.75pt
    \pgfmathsetmacro{\ratio}{100*(\val+1)/0.25}
    \cellcolor{strongred!\ratio!softred}#2%
  \else\ifdim\val pt<-0.5pt
    \pgfmathsetmacro{\ratio}{100*(\val+0.75)/0.25}
    \cellcolor{softred!\ratio!lightred}#2%
  \else\ifdim\val pt<-0.25pt
    \pgfmathsetmacro{\ratio}{100*(\val+0.5)/0.25}
    \cellcolor{lightred!\ratio!verylightred}#2%
  \else\ifdim\val pt<0pt
    \pgfmathsetmacro{\ratio}{100*(\val+0.25)/0.25}
    \cellcolor{verylightred!\ratio!neutral}#2%
  \else\ifdim\val pt<0.25pt
    \pgfmathsetmacro{\ratio}{100*\val/0.25}
    \cellcolor{neutral!\ratio!verylightgreen}#2%
  \else\ifdim\val pt<0.5pt
    \pgfmathsetmacro{\ratio}{100*(\val-0.25)/0.25}
    \cellcolor{verylightgreen!\ratio!lightgreen}#2%
  \else\ifdim\val pt<0.75pt
    \pgfmathsetmacro{\ratio}{100*(\val-0.5)/0.25}
    \cellcolor{lightgreen!\ratio!softgreen}#2%
  \else
    \pgfmathsetmacro{\ratio}{100*(\val-0.75)/0.25}
    \cellcolor{softgreen!\ratio!stronggreen}#2%
  \fi\fi\fi\fi\fi\fi\fi
}

\newcommand{\smoothgradient}[2]{%
  { 
    \pgfmathsetmacro{\val}{#1}%
    \ifdim\val pt<0pt
        \pgfmathsetmacro{\ratio}{min(100,-100*\val)}%
        \pgfmathparse{int(247-32 *\ratio/100)}\edef\r{\pgfmathresult}%
        \pgfmathparse{int(247-199*\ratio/100)}\edef\g{\pgfmathresult}%
        \pgfmathparse{int(247-208*\ratio/100)}\edef\b{\pgfmathresult}%
    \else
        \pgfmathsetmacro{\ratio}{min(100,100*\val)}%
        \pgfmathparse{int(247-220*\ratio/100)}\edef\r{\pgfmathresult}%
        \pgfmathparse{int(247-127*\ratio/100)}\edef\g{\pgfmathresult}%
        \pgfmathparse{int(247-192*\ratio/100)}\edef\b{\pgfmathresult}%
    \fi
    \cellcolor[RGB]{\r,\g,\b}#2%
  } 
}

\definecolor{greencell}{HTML}{2e7d43}
\definecolor{redcell}{HTML}{FF3333}

\Crefname{section}{Section}{Sections}
\Crefname{subsection}{Subsection}{Subsections}
\Crefname{appendix}{Appendix}{Appendices}
\Crefname{figure}{Figure}{Figures}
\Crefname{table}{Table}{Tables}

\newtcolorbox{instructionsbox}[1][]{
  breakable,
  colframe=cyan!75!black,    
  colback=green!5!white,     
  coltitle=black,            
  title=#1,                  
  rounded corners,           
  boxrule=0.5mm,             
  boxsep=5pt,                
  toptitle=1mm,              
  bottomtitle=1mm,           
  left=10pt,                 
  right=10pt,                
  top=5pt,                   
  bottom=5pt,                
  fonttitle=\bfseries        
}

\newtcolorbox{promptbox}[1][]{
  breakable,
  colframe=orange!60!brown,    
  colback=brown!10!white,     
  coltitle=black,            
  title=#1,                  
  rounded corners,           
  boxrule=0.5mm,             
  boxsep=5pt,                
  toptitle=1mm,              
  bottomtitle=1mm,           
  left=10pt,                 
  right=10pt,                
  top=5pt,                   
  bottom=5pt,                
  fonttitle=\bfseries        
}

\title{Economies of Open Intelligence:\\Tracing Power \& Participation in the Model Ecosystem}

\usepackage{authblk}

\makeatletter
\renewcommand\AB@affilsepx{\hspace{1em}\protect\Affilfont}
\makeatother

\newcommand{\eqmark}{\textsuperscript{\dag}}

\author[1,2]{Shayne Longpre\eqmark}
\author[2,3]{Christopher Akiki}
\author[2,4]{Campbell Lund}
\author[2,5]{Atharva Kulkarni}
\author[2,6]{Emily~Chen}
\author[7]{Irene Solaiman}
\author[7]{Avijit Ghosh}
\author[7]{Yacine Jernite}
\author[7]{Lucie{-}Aimée Kaffee\eqmark}

\affil[1]{MIT}
\affil[2]{Data Provenance Initiative}
\affil[3]{ScaDS.AI Leipzig}
\affil[4]{University of Edinburgh}
\affil[5]{University~of~Southern~California} 
\affil[6]{UNC at Chapel Hill}
\affil[7]{Hugging Face}

\date{}

\begin{document}

\maketitle

\begingroup
\renewcommand\thefootnote{\fnsymbol{footnote}}
\setcounter{footnote}{0}
\footnotetext[2]{Equal contribution. Correspondence: slongpre@media.mit.edu}
\endgroup

\doparttoc
\faketableofcontents

\begin{abstract}
Since 2019, the Hugging Face Model Hub has been the primary global platform for sharing open weight AI models. 
By releasing a dataset of the complete history of weekly model downloads (June 2020–August 2025) alongside model metadata, we provide the most rigorous examination to-date of concentration dynamics and evolving characteristics in the open model economy. 
Our analysis spans 851,000 models, over 200 aggregated attributes per model, and 2.2B downloads. 
We document a fundamental rebalancing of economic power: US open-weight industry dominance by Google, Meta, and OpenAI has declined sharply in favor of unaffiliated developers, community organizations, and, as of 2025, Chinese industry, with DeepSeek and Qwen models potentially heralding a new consolidation of market power. 
We identify statistically significant shifts in model properties---a 17$\times$ increase in average model size, rapid growth in multimodal generation (3.4$\times$), quantization (5$\times$), and mixture-of-experts architectures (7$\times$)---alongside concerning declines in data transparency, with open weights models surpassing truly open source models for the first time in 2025. 
We expose a new layer of developer intermediaries that has emerged, focused on quantizing and adapting base models for both efficiency and artistic expression. 
To enable continued research and oversight, we release the complete dataset with an interactive dashboard for real-time monitoring of concentration dynamics and evolving properties in the open model economy.
\end{abstract}

\begin{center}
\vspace{-2mm}
\begin{tabular}{@{}l@{\hspace{0.7em}}l@{\hspace{0.7em}}l@{}}
{\twemoji[width=1em]{trophy}} & \textbf{Dashboard} & \href{https://huggingface.co/spaces/economies-open-ai/open-model-evolution}{huggingface.co/spaces/economies-open-ai/open-model-evolution}
\end{tabular}
\end{center}

\section{Introduction}

The concentration of power in artificial intelligence across computational resources \citep{lehdonvirta2024compute}, training data \citep{buolamwini2018gender,longpre2023data,longprebridging}, and model development, has emerged as a central focus for the fairness, safety, and control of the AI economy \citep{hopkins2025ai}. 
Understanding where power concentrates, how it shifts over time, and which actors control the development and distribution of AI models is essential for effective oversight and equitable access \citep{crawford2021atlas, noble2018algorithms}. 
While prior work has examined the broader AI supply chain \citep{hopkins2025ai}, the topology of Hugging Face model relationships \citep{laufer2025anatomy,horwitz2025we}, or the implications of economic consolidation in the \emph{closed} model AI market \citep{korinek2025concentrating, vipra2023concentration, vipra2023computational, nagle2025latent}, no study has systematically traced how economic power concentrates and diffuses in the \emph{open} model ecosystem over time, nor rigorously examined which model characteristics are waxing or waning in adoption.

The Hugging Face Model Hub provides a unique window into these dynamics. What began as a mode of distributing the open PyTorch formats of BERT and GPT-2 models in 2019 has evolved into the primary global platform for sharing open weight AI models, now hosting 2M+ models with 1.7B unique, cumulative downloads.\footnote{Note: we use a more precise measure of downloads than the raw counts reported on the Hub (see \Cref{sec:methodology}).} This platform has become central to the international adoption and distribution of open weights models for both research and production use, spanning the full range of general-purpose AI tasks, modalities (text, speech, image, video, tabular), and languages. 
As such, it offers the most comprehensive view available into the evolution of the open AI model ecosystem—especially the dynamics of control, consolidation, and market power.

In this work, we aggregate and publicly release the complete historical download logs for the Hugging Face Model Hub spanning June 2020 to August 2025, merged with extensive metadata on model training methods, architectures, modalities, languages, developer country of origin, documentation quality, and access restrictions. This represents the largest and most rigorous study of open model usage to date, encompassing detailed analysis of 851,000 models (97.6\% of all downloads) with additional collected metadata, often absent from automated records. 
Using temporal download patterns as the closets available proxy for model adoption, we trace long-term shifts and measure the concentration of market power across models, developers, and countries using established economic metrics including the Herfindahl-Hirschman Index (HHI) and Gini coefficient.

Our analysis reveals several critical findings for the open AI economy:
\begin{enumerate}
    \item \textbf{Declining US industry dominance, rising influence of China and unaffiliated developers}: Concentration measures show a steep decline in US industry's market share by Google, Meta, and OpenAI, starting in 2022. Power has diffused to unaffiliated users, community developers, and recently to Chinese industry, with DeepSeek and Qwen models potentially ushering in a new consolidation of market power led by Chinese developers.
    
    \item \textbf{Shift toward larger, multimodal, and computationally efficient architectures}: Models downloaded in 2025 grew to 17$\times$ the average 2020 parameter count, and exhibited more multimodal and video generation capabilities (3.4$\times$ increase). To accommodate these larger sizes, we see rising use of quantization techniques (5$\times$), parameter-efficient finetuning (1.4$\times$), and mixture-of-experts architectures (7$\times$).
    
    \item \textbf{Emergence of a developer intermediary layer}: Organizations that quantize, finetune, re-package, or build artistic adapters for major base models have surged in popularity, constructing a fundamental new layer of model intermediaries between base model creators and end users.
    
    \item \textbf{Sharp decline in open source compliance and data transparency}: Data transparency has deteriorated significantly, with the proportion of downloads for models disclosing available training data falling from 79.3\% (2022) to just 39\% (2025). For the first time in 2025, downloads of open weights models (lacking training data disclosure) surpassed truly open source models that meet the requirements of the Open Source Initiative definition\footnote{\url{https://opensource.org/ai/open-source-ai-definition}}.
    
    \item \textbf{Public release of data and live monitoring dashboard:}\footnote{Dashboard is available here: \url{https://huggingface.co/spaces/mmpr/open-model-evolution}. The live public dashboard gives slightly different results as it uses publicly available downloads data, which is less precise than the internal, deduplicated data used in this paper.} To enable continued research, transparency, and oversight, we release the complete dataset alongside an interactive dashboard for real-time monitoring of concentration dynamics, market power shifts, and evolving model properties in the open AI ecosystem.
\end{enumerate}

These findings provide critical empirical grounding for policy discussions around AI governance, economic concentration, and the preservation of open and equitable access to AI capabilities.
\vspace{-2mm}
\section{Experimental Methodology} 
\label{sec:methodology}
\vspace{-2mm}

\begin{figure}
    \centering
    \includegraphics[width=\linewidth]{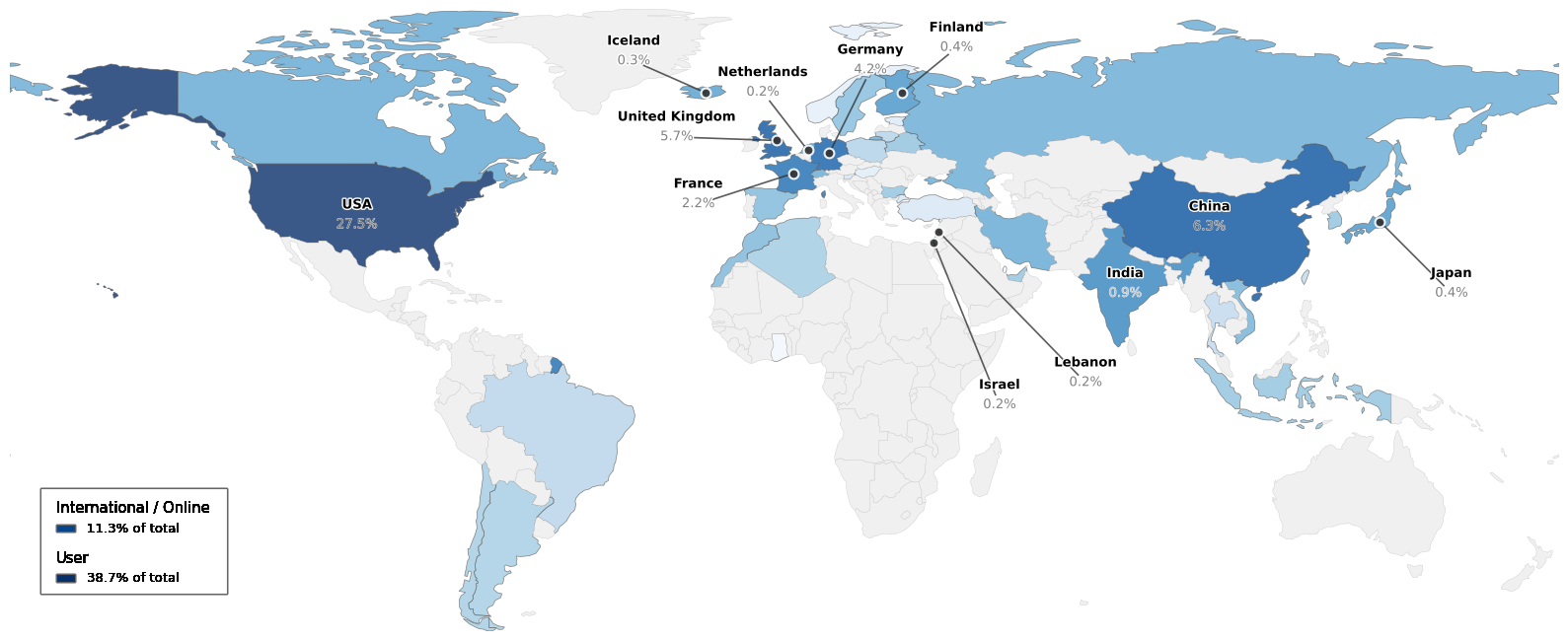}
    
    \vspace{0.5em} 
    \small
\setlength{\tabcolsep}{3pt}
\begin{adjustbox}{width=\textwidth}
\begin{tabular}{l|lrl|lrl|lrl|lrl}
\toprule
  & \multicolumn{3}{l|}{\textsc{Top Countries}} & \multicolumn{3}{l|}{\textsc{Top Developers}} & \multicolumn{3}{l|}{\textsc{Top Models}} \\
\midrule
\multirow{10}{*}{\rotatebox{90}{\textsc{\textbf{\emph{All-time}}}}}
& Unaffil. User & {\twemoji{1f464}} & 38.7\% & Google & {\twemoji{flag: United States}} \includegraphics[height=0.25cm]{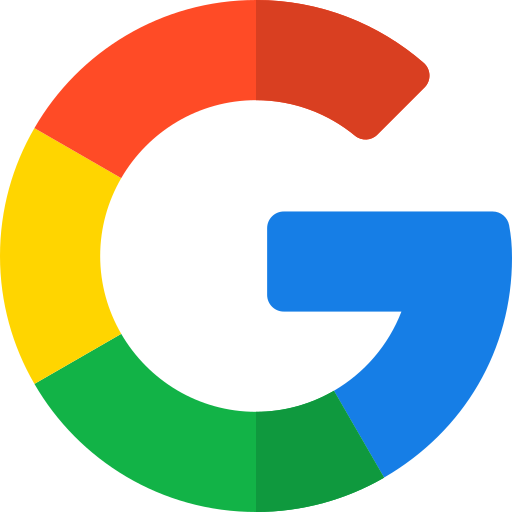} & 6.5\% & adetailer &  {\twemoji{1f464}} {\twemoji{1f9e9}} & 2.9\% \\ 
& USA & {\twemoji{flag: United States}} & 27.5\% & stable-diffusion & {\twemoji{flag: United Kingdom}} \includegraphics[height=0.25cm]{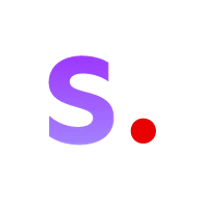} & 5.4\% & bert-base-uncased & {\twemoji{flag: United States}} \includegraphics[height=0.25cm]{custom_emojis/google.png} {\twemoji{1f9e9}} & 2.9\% \\ 
& Internat./Online & {\twemoji{1f310}} & 11.3\% & Bingsu &  {\twemoji{1f464}} & 5.3\% & yolo-world-mirror &  {\twemoji{1f464}} {\twemoji{1f9e9}} & 2.4\% \\ 
& China & {\twemoji{flag: China}} & 6.3\% & lllyasviel &  {\twemoji{1f464}} & 4.7\% & clip-vit-large-patch14 & {\twemoji{flag: United States}} \includegraphics[height=0.25cm]{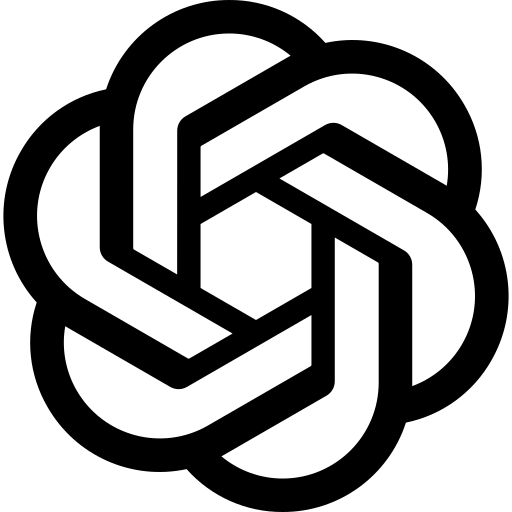} {\twemoji{1f9e9}} & 2.1\% \\ 
& UK & {\twemoji{flag: United Kingdom}} & 5.7\% & Facebook & {\twemoji{flag: United States}} \includegraphics[height=0.25cm]{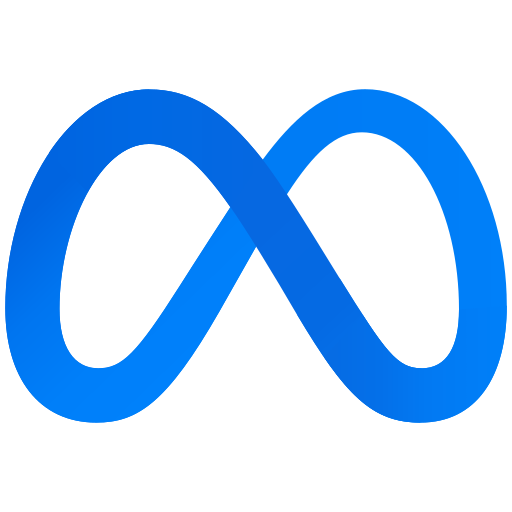} & 4.6\% & stable-diffusion-xl-base-1.0 & {\twemoji{flag: United Kingdom}} \includegraphics[height=0.25cm]{custom_emojis/stable-diffusion.png} {\twemoji{1f5bc}} & 1.4\% \\ 
& Germany & {\twemoji{flag: Germany}} & 4.2\% & lmstudio-community & {\twemoji{1f310}} \includegraphics[height=0.25cm]{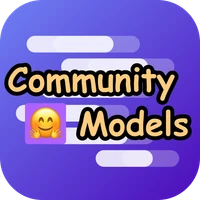} & 4.0\% & distilbert-base-uncased & {\twemoji{flag: United States}} \includegraphics[height=0.25cm]{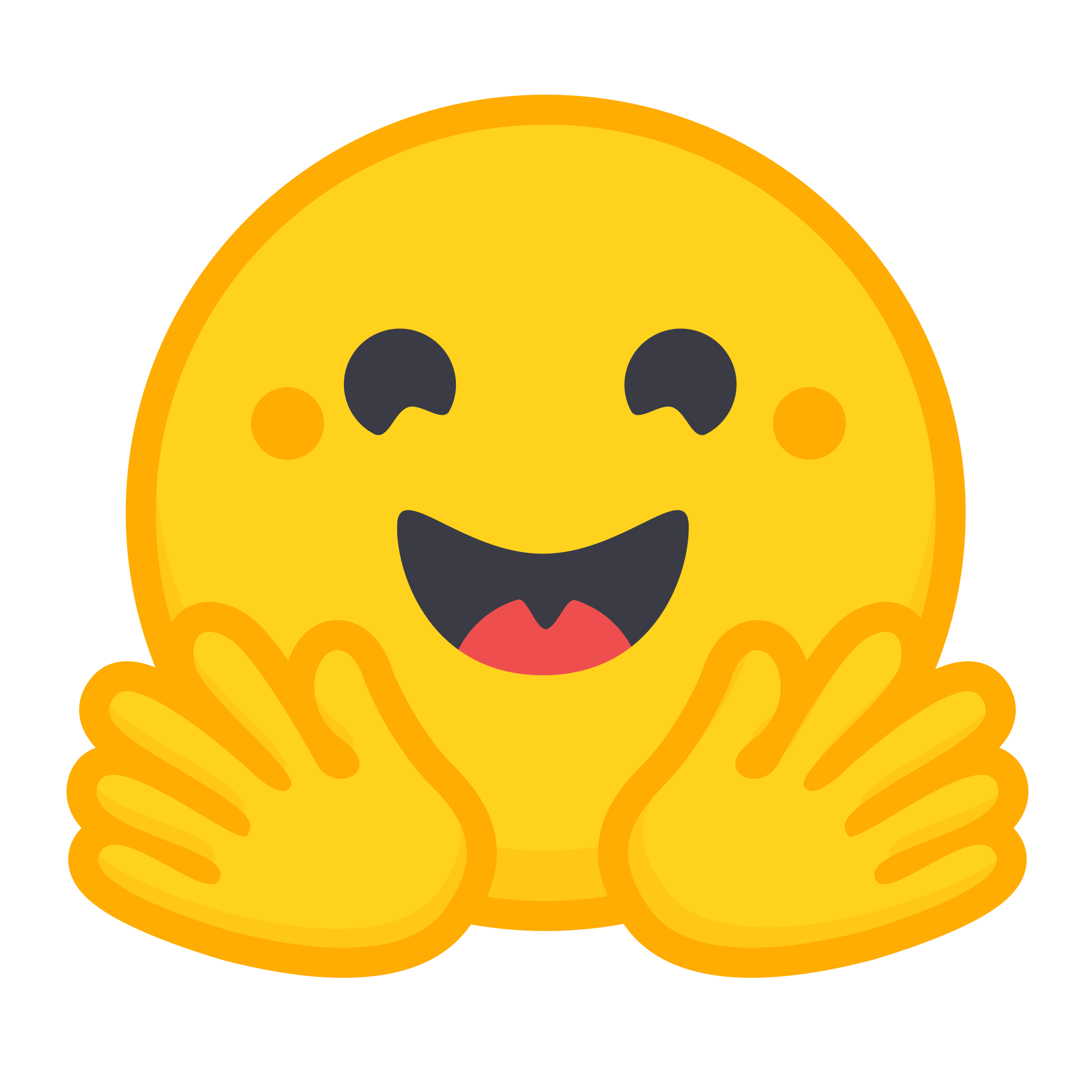} {\twemoji{1f9e9}} & 1.1\% \\ 
& France & {\twemoji{flag: France}} & 2.2\% & OpenAI & {\twemoji{flag: United States}} \includegraphics[height=0.25cm]{custom_emojis/open-ai.png} & 3.6\% & fav\_models &  {\twemoji{1f464}} {\twemoji{1f539}} & 0.9\% \\ 
& India & {\twemoji{flag: India}} & 0.9\% & sd-concepts-library & {\twemoji{1f310}} {\twemoji{1f3e2}} & 2.8\% & misc &  {\twemoji{1f464}} {\twemoji{1f539}} & 0.9\% \\ 
& Finland & {\twemoji{flag: Finland}} & 0.4\% & timm & {\twemoji{flag: United States}} \includegraphics[height=0.25cm]{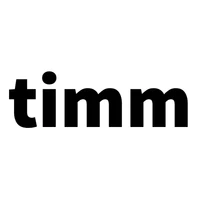} & 2.3\% & dependencies &  {\twemoji{1f464}} {\twemoji{1f539}} & 0.8\% \\ 
& Japan & {\twemoji{flag: Japan}} & 0.4\% & deepseek-ai & {\twemoji{flag: China}} \includegraphics[height=0.25cm]{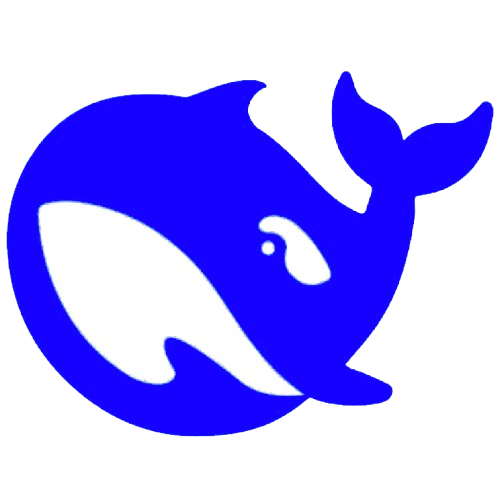} & 2.3\% & stable-diffusion & {\twemoji{flag: Germany}} {\twemoji{1f3e2}} {\twemoji{1f5bc}} & 0.8\% \\ 
\midrule
\multirow{10}{*}{\rotatebox{90}{\textsc{\textbf{\emph{Aug. 2024 - Aug. 2025}}}}}
& Unaffil. User & {\twemoji{1f464}} & 33.8\% & lmstudio-community & {\twemoji{1f310}} \includegraphics[height=0.25cm]{custom_emojis/lmstudio-community.png} & 16.4\% & DeepSeek-R1 & {\twemoji{flag: China}} \includegraphics[height=0.25cm]{custom_emojis/deepseek-ai.png} {\twemoji{1f4dd}} & 2.8\% \\ 
& Internat./Online & {\twemoji{1f310}} & 23.8\% & deepseek-ai & {\twemoji{flag: China}} \includegraphics[height=0.25cm]{custom_emojis/deepseek-ai.png} & 9.6\% & stable-diffusion-v1-5 & {\twemoji{flag: United Kingdom}} \includegraphics[height=0.25cm]{custom_emojis/stable-diffusion.png} {\twemoji{1f5bc}} & 2.4\% \\ 
& China & {\twemoji{flag: China}} & 17.1\% & comfy & {\twemoji{flag: United States}} \includegraphics[height=0.25cm]{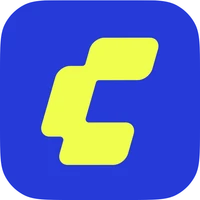} & 5.4\% & flux\_text\_encoders & {\twemoji{flag: United States}} \includegraphics[height=0.25cm]{custom_emojis/comfy.png} {\twemoji{1f9e9}} & 1.2\% \\ 
& USA & {\twemoji{flag: United States}} & 15.8\% & Qwen & {\twemoji{flag: China}} \includegraphics[height=0.25cm]{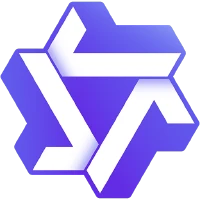} & 4.6\% & Wan\_2.1\_ComfyUI\_repackaged & {\twemoji{flag: United States}} \includegraphics[height=0.25cm]{custom_emojis/comfy.png} {\twemoji{1f5bc}} & 1.2\% \\ 
& UK & {\twemoji{flag: United Kingdom}} & 3.6\% & stable-diffusion & {\twemoji{flag: United Kingdom}} \includegraphics[height=0.25cm]{custom_emojis/stable-diffusion.png} & 3.6\% & stable-diffusion-v1-5-archive & {\twemoji{flag: United States}} \includegraphics[height=0.25cm]{custom_emojis/comfy.png} {\twemoji{1f5bc}} & 1.1\% \\ 
& India & {\twemoji{flag: India}} & 3.4\% & strangerzonehf & {\twemoji{flag: India}} \includegraphics[height=0.25cm]{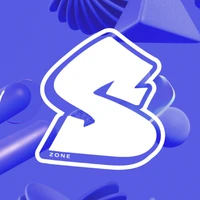} & 3.4\% & Llama-3.2-1B-Instruct-q4f16\_1-MLC & {\twemoji{1f310}} \includegraphics[height=0.25cm]{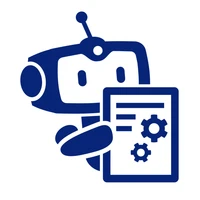} {\twemoji{1f4dd}} & 1.0\% \\ 
& France & {\twemoji{flag: France}} & 1.0\% & alvdansen &  {\twemoji{1f464}} & 3.1\% & DeepSeek-V3 & {\twemoji{flag: China}} \includegraphics[height=0.25cm]{custom_emojis/deepseek-ai.png} {\twemoji{1f4dd}} & 1.0\% \\ 
& Germany & {\twemoji{flag: Germany}} & 0.4\% & mlx-community & {\twemoji{1f310}} \includegraphics[height=0.25cm]{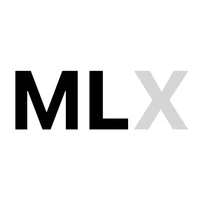} & 2.6\% & Qwen2.5-0.5B-Instruct-q4f16\_1-MLC & {\twemoji{1f310}} \includegraphics[height=0.25cm]{custom_emojis/mlc-ai.png} {\twemoji{1f4dd}} & 0.9\% \\ 
& Singapore & {\twemoji{flag: Singapore}} & 0.2\% & Shakker-Labs & {\twemoji{flag: United States}} {\includegraphics[height=0.25cm]{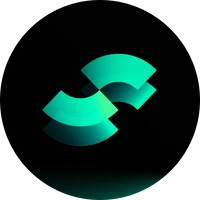}} & 2.5\% & clip-vit-base-patch32-ONNX & {\twemoji{1f310}} \includegraphics[height=0.25cm]{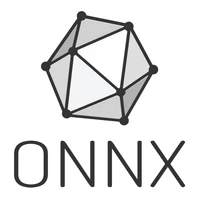} {\twemoji{1f9e9}} & 0.8\% \\ 
& Switzerland & {\twemoji{flag: Switzerland}} & 0.2\% & openfree &  {\twemoji{1f464}} & 2.2\% & Janus-Pro-7B & {\twemoji{flag: China}} \includegraphics[height=0.25cm]{custom_emojis/deepseek-ai.png} {\twemoji{1f4dd}} & 0.8\% \\ 
\bottomrule
\end{tabular}
\end{adjustbox}
\vspace{-1mm}
\caption{
\textbf{Top:} The \textsc{Top 12 Nation Map} ranked by the all-time \includegraphics[height=0.25cm]{custom_emojis/hugging-face.png} downloads of their models. \newline
\textbf{Bottom:} The top 10 \includegraphics[height=0.25cm]{custom_emojis/hugging-face.png} downloads \textsc{Leaderboard} for countries, developers, and models, with their download percentages. 
Both the map and the leaderboard use \rollingwindowfilter{} to mitigate inauthentic downloads.
The \textbf{\emph{All-time}} section reflects all time downloads, whereas the \textbf{\emph{Aug. 2024 - Aug. 2025}} reflects all downloads for models created within the last year (August 2024 to August 2025).
Symbols indicate details about the model: {\twemoji{1f9e9}} = embedding and classification models; {\twemoji{1f4dd}} = text generation, {\twemoji{1f5bc}} = image generation, {\twemoji{1f399}} = speech generation, {\twemoji{1f3a5}} = video generation, {\twemoji{1f310}} = international/online organization, {\twemoji{1f464}} = unaffiliated user.
}
\label{tab:leaderboard}
    
    \label{fig:national-concentration}
    \vspace{-5mm}
\end{figure}

Our temporal usage analysis of the open model ecosystem is made feasible by merging metadata from several sources: Hugging Face's entire history of model downloads, automatic crawls of the Model Hub directories, and manually collected annotations: the derivation tree from other models, architectures, input/output modalities, training/inference methods, languages, transparency, documentation, and access restrictions. 
Full data collection details are provided in \Cref{app:dataset-details}.

\textbf{Usage Data.} For the usage data, we aggregate model downloads, at a weekly cadence, from the Hugging Face Model platform, since they first started recording in June 2020.
We only count one unique download per user per day, and filter out models with $<200$ total downloads, to not over-count repetitive automatic processes, or models that are not frequently used beyond the developer.
These filters yield 851k models of all 1.88M available on the Hub, however, this sample accounts for 97.6\% of all downloads.
We denote this as the \modelpopulation.
From this population, we conduct a stratified sample of the top $200$ most downloaded models for each of the 265 weeks between June 2020 and August 2025.
This yields $2875$ models, comprising $49.6\%$ of all downloads, which we call the \headsample.
We collect significantly more detailed metadata using trained human annotators about the model sizes, languages, derivations, architectures, training methods, and data sources, which Hugging Face metadata often misses.

Note that there are few sources for open model \emph{usage}, and few are systematically reliable.
For instance, we investigated the well-known OpenRouter, however it's model selection and user-base is limited, and the metrics are skewed towards models that do not have APIs available elsewhere.\footnote{\url{https://openrouter.ai/rankings}}
We posit that Hugging Face downloads (after we have normalized and filtered) offers by far the most reliable metric, indicative of realistic measures of community adoption.
Even so, raw counts show a disproportionate number of downloads for pre-2023 models.
But these downloads are driven by outdated automatic software, that continuously loads models without any meaningful usage.\footnote{For instance vLLM automatically downloads OPT-125M as part of its CI/CD despite minimal actual usage (see \url{https://github.com/vllm-project/vllm/issues/7053}).}
This can lead to skewed estimates of usage and popularity.
To account for this, we use a ``\rollingwindowfilterbf'' metric, where downloads are only counted if the model was created within a year of the download.
We posit this filter isolates more authentic popularity and usage trends, before the download signal is overcome by automatic processes that don't actually use the models.
Typically, more performant or efficient versions of models are available well within one year.

Lastly, there is a choice on how to attribute credit when a model A is downloaded, but it was adapted (finetuned, quantized, etc) from a base model B.
Both choices are valid though they offer different assumptions.
For some analyses we use the latter, which we refer to as ``\derivedAuthorbf'', applying credit to the base models (and their developers).

\textbf{Model Characteristic Data.}
To ascertain trends in model usage, we also collect significant metadata that is not already available for models on the Hugging Face Hub.
We curated detailed instructions, and hired expert annotators, compensated at $25\$$ per hour, with regular content reviews by the authors, to ensure 90\%+ annotation accuracy.
To ascertain model sizes, we use the Model Hub \texttt{safetensors} field where available, and otherwise estimate the parameter count from the models' file sizes.
We employed RANSAC (Random Sample Consensus) regression, a robust estimator that automatically identifies and excludes outliers arising from varying model architectures and compression ratios.
This method obtained a validation $R^2=0.86$, and increased model size coverage from 23.6\% to 98\% of models.
Full details of the annotation process are provided in \Cref{app:annotation-taxonomy-details}.

\textbf{Economic Measures of Market Concentration.}
The field of economics has developed concentration metrics to measure the distributional inequality in economic systems. 
Most widely recognized is the Herfindahl-Hirschman Index (HHI).
Calculated as the sum of squared market shares, it ranges from near zero (perfect competition) to 10,000 (monopoly)---though we squash this from 0 to 1 \cite{hirschman1964paternity, rhoades1993herfindahl}. 
For a complementary perspective, the Gini coefficient, ranging from 0 (perfect equality) to 1 (maximum inequality), measures the cumulative deviation from perfect equality across the entire distribution \cite{gini1912variability, sen1997economic}. 
While both metrics increase with concentration, they capture distinct phenomena: HHI emphasizes market leaders at the head of the distribution, whereas the Gini coefficient evaluates the equality across all participants equally \cite{kwoka1985herfindahl}.

\begin{figure*}[t]             
  \centering
  
  \begin{subfigure}{\textwidth}
    \includegraphics[width=\textwidth,
                     clip,trim=0 2 0 0]{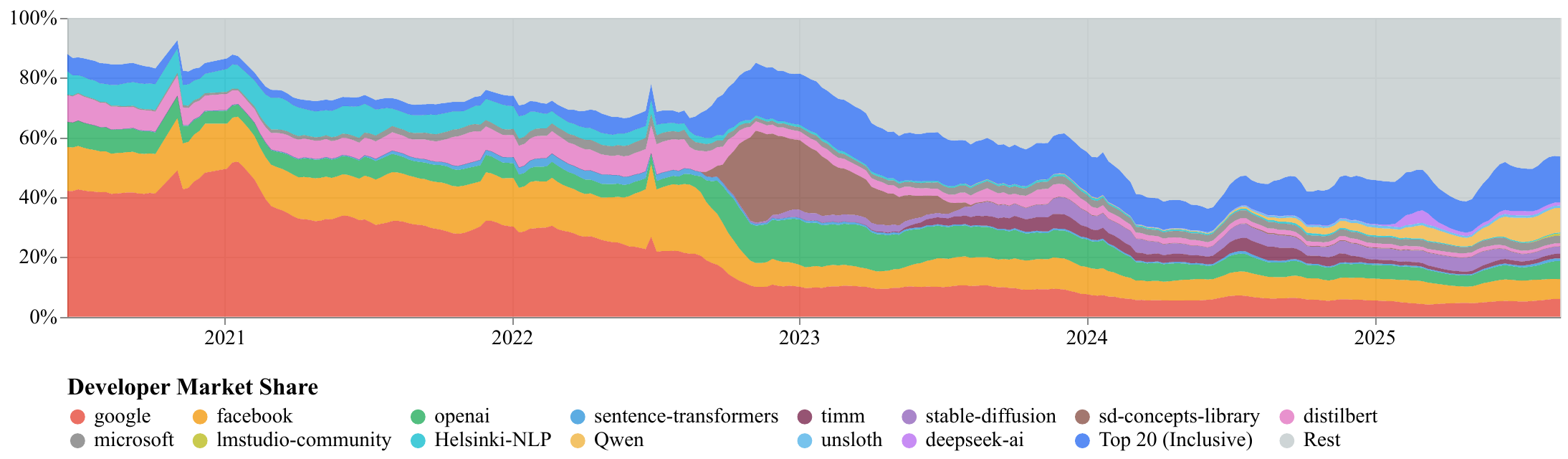}
  \end{subfigure}

  \vspace{-3pt}
  
  \begin{subfigure}{\textwidth}
    \includegraphics[width=\textwidth,
                     clip,trim=0 2 0 0]{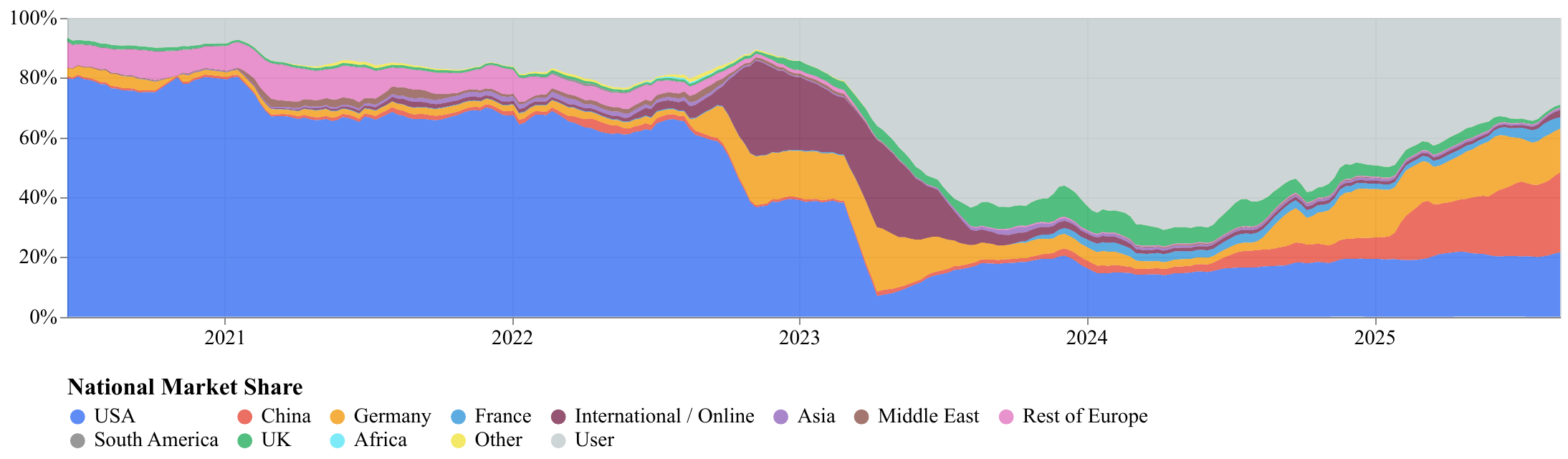}
  \end{subfigure}
  
  \caption{
  \textbf{Top:} Developer Download Share over time, using the \rollingwindowfilter{} and applying \derivedAuthor{}. Where Google, Meta, and OpenAI once dominated market share (2021-2024), their influence has subsided for other developers beyond the Top 20 have gradually increased from 20\% to >50\% share.
  \textbf{Bottom:} National Download Share over time, using the \rollingwindowfilter{} and applying \derivedAuthor{}.
  Where the US and Europe once dominated market share (2021-2023), now Users, China, and Germany have become prominent contributors.
  Both plots use a 1-year \rollingwindowfilter{} to better estimate authentic usage.
    }
  \label{fig:temporal_rankings}
  \vspace{-4mm}
\end{figure*}

\begin{figure*}[t]             
  \centering
  
  \begin{subfigure}{\textwidth}
    \includegraphics[width=\textwidth,
                     clip,trim=0 2 0 0]{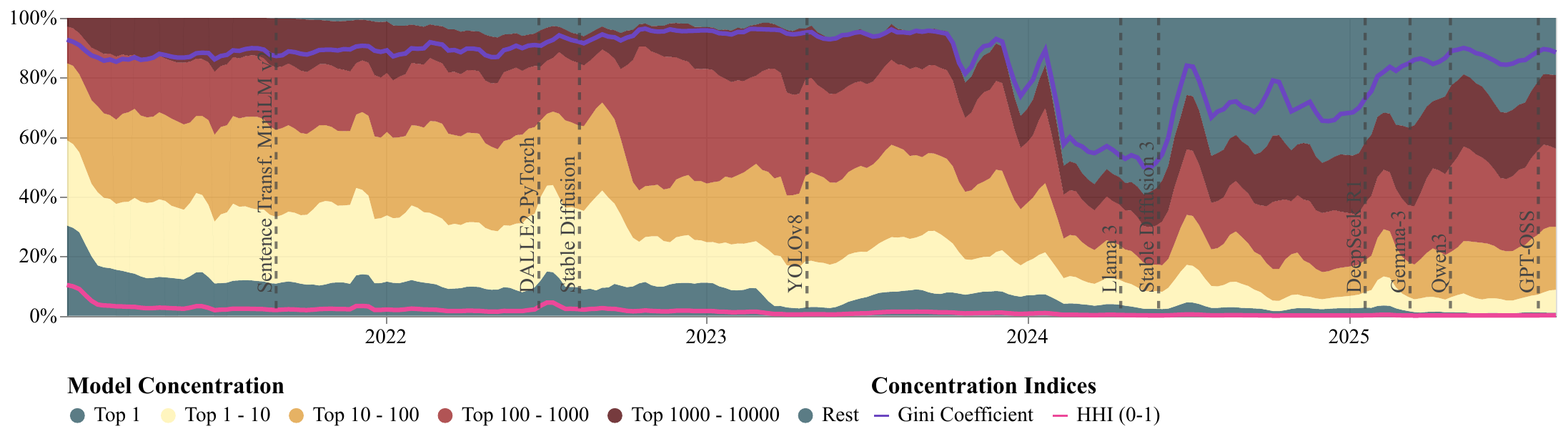}
  \end{subfigure}
  
  \vspace{-3pt}
  
  \begin{subfigure}{\textwidth}
    \includegraphics[width=\textwidth,
                     clip,trim=0 2 0 0]{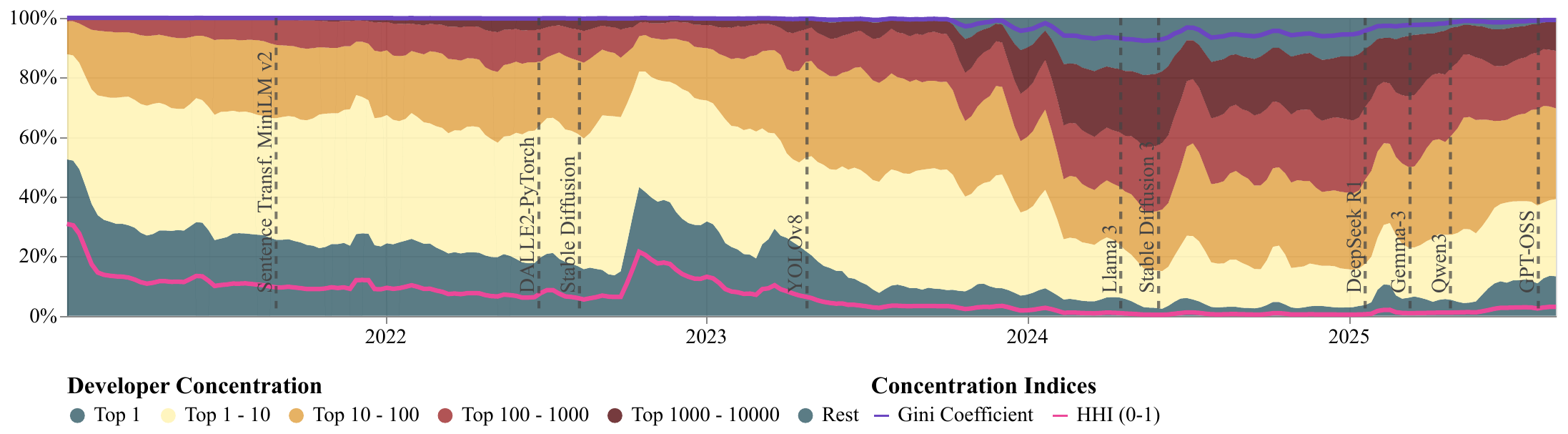}
  \end{subfigure}

  \vspace{-3pt}
  
  \begin{subfigure}{\textwidth}
    \includegraphics[width=\textwidth,
                     clip,trim=0 2 0 0]{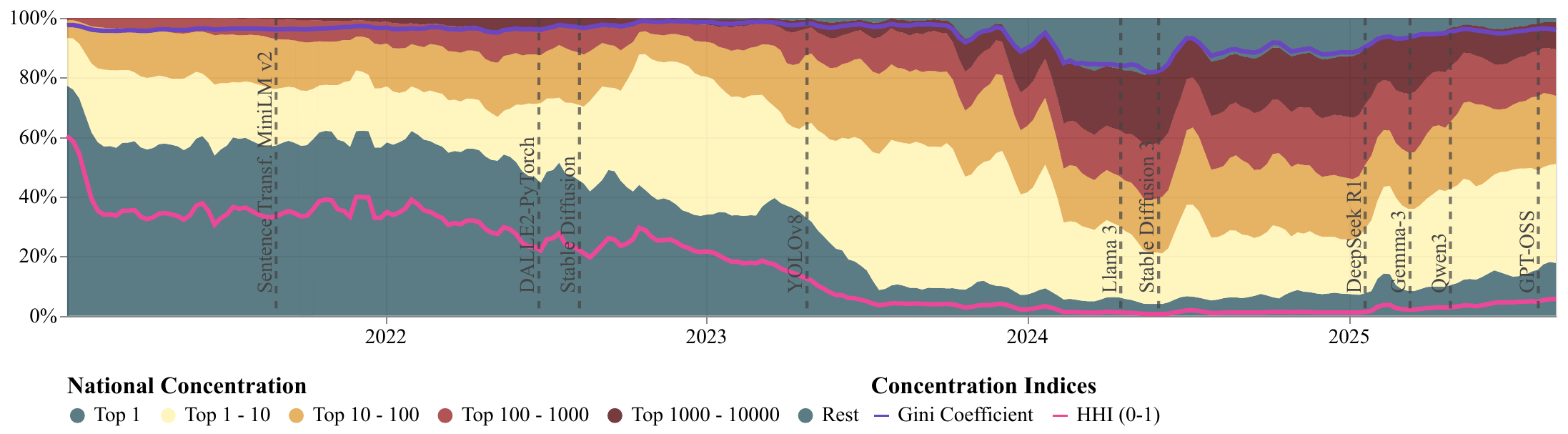}
  \end{subfigure}
  
  \caption{  
    \textbf{Top:} Model economic concentration over time.  
    \textbf{Middle:} Developer economic concentration over time.
    \textbf{Bottom:} National economic concentration over time.
    In each plot we measure the share of downloads allocated to ranked segments of the open model economy.
    Economic measures of concentration are also displayed in purple (the Gini coefficient) and pink (the Hherfindahl-Hirschman Index from 0-1).
    \textbf{Across levels of abstraction (model, developer, nation) economic concentration first declined significantly, but has started to rise again in 2025.}
    }
  \label{fig:temporal_concentration}
  \vspace{-4mm}
\end{figure*}

\vspace{-2mm}
\section{Concentration of Power: Open Models, Developers, \& Nations}
\label{sec:concentration}
\vspace{-2mm}

To gain an insight into the concentration of power in the open model ecosystem, we trace market share at the model, developer, and national market-levels.
Where prior work has mapped the concentration of computational resources \citep{lehdonvirta2024compute}, the concentration and geographical representation of training data \citep{longpre2023data,longprebridging}, and assessed static views of the Hugging Face model graph \citep{laufer2025anatomy, horwitz2025we}, this is the first work to comprehensively map open model \emph{usage} over the full historical time range that Hugging Face has recorded.
This enables us to trace patterns of economic concentration, and their evolution.
We first examine the all-time snapshot of the ecosystem (upper \Cref{tab:leaderboard}), then explore three \emph{periods} of model development---traced in the temporal breakdowns of market share (\Cref{fig:temporal_rankings}) and economic concentration (\Cref{fig:temporal_concentration}).

\textbf{An All-Time Snapshot of the Ecosystem.}
First, in \Cref{tab:leaderboard} we take a snapshot of the current open model ecosystem, from two perspectives: \textbf{\emph{All-time}} and authentic usage from the \textbf{\emph{Aug 2024 - Aug 2025}}.
The global map and \textbf{\emph{All-time}} segment of the \textsc{Leaderboard} show download shares from June 2020 - Sept 2025, filtered by the \rollingwindowfilter{} to mitigate inauthentic usage.
The interactive dashboard allows readers to see leaderboard statistics both with and without these filters.

Across all-time we see the USA, Western Europe (UK, Germany, France), have dominated development, through mostly industry organizations: Google, OpenAI, Meta, HuggingFace's Sentence-Transformers, and Stable-Diffusion.
American text and image embedding/classification models, based on CLIP \citep{pmlr-v139-radford21a}, BERT \citep{devlin-etal-2019-bert}, YOLO \citep{Redmon_2016_CVPR}, and their variants account for many of the all-time downloads.

\textbf{The Foundation Embeddings Era (before late 2022)} 
Prior to late 2022, the open model ecosystem was characterized by high market concentration and geographic consolidation. 
The top three US-based organizations—Google, Meta, and OpenAI—commanded between 40-60\% of cumulative downloads, while the USA alone represented over 60 of national market share. 
\Cref{fig:temporal_concentration} shows that in this period, the top 100 developers comprised over 90\% of all downloads.
HHI measures at the national and developer level were highest during this period, beginning around 0.3 and 0.6 in 2021.
This era was dominated by a homogeneous consolidation around certain types of models: encoder-based architectures, with text encoder-only transformers, and embedding/classification heads comprising 76.8\% and 75.2\% of all downloads, respectively (see \Cref{tab:temporal-shifts}). 
Models were relatively small (average 217M parameters) with high transparency standards: 79.3\% disclosed their training data availability. 
The technological focus centered on foundational capabilities like BERT \citep{devlin-etal-2019-bert}, CLIP \citep{pmlr-v139-radford21a}, DistilBERT \citep{sanh2020distilbertdistilledversionbert}, and sentence transformers \citep{reimers-gurevych-2019-sentence} that would later enable more sophisticated applications.

\textbf{The Generative Diffusion Period (late 2022 to early 2024)}
This period witnessed a dramatic democratization of AI development following Stable Diffusion's release \citep{Rombach_2022_CVPR}---a \emph{diffusion} both of generative architectures and of developer participation.
Diffusion-based networks spike to 20\% of all downloads, compared to $<1\%$ a year prior.
This sparked the other defining feature of this period: market concentration plummeted as international and online organizations, followed by unaffiliated users, began developing on these models.
Entities like CompVis, sd-concepts-library, and communities that develop text and image adapters \citep{pmlr-v97-houlsby19a} surged to prominence, collectively capturing notable portions of the download share (\Cref{fig:temporal_rankings}). 
Many of these user communities began to form around libraries of artistic styles and renditions that could be applied to Stable Diffusion models.

\Cref{fig:temporal_concentration} shows this period is marked predominantly by a diffusion of market power.
Measures of economic concentration, especially HHI and the Top-10 share, begin declining precipitously, especially in early 2024.
Notably, the ``International/Online'' category in national share experienced its largest spike, and the "Rest" category of developers grew from ~20\% to over 50\% of market share. 
Simultaneously, the popularity of OpenAI models rose significantly with wide interest in their image embedding and speech generation models, CLIP-VIT-Large \citep{pmlr-v139-radford21a}, and Whisper \citep{pmlr-v202-radford23a}.
This era marked a fundamental shift from industry-led development to grassroots innovation, as thousands of community members created LoRA adapters, textual inversions, and fine-tuned variants. 
The barriers to AI development lowered substantially, enabling individual users and small collectives to meaningfully participate in the ecosystem.

\textbf{The Sino-Multimodal Period (late 2024 to the present)}
The most recent period represents a plurality of changes: a fundamental re-balancing of global power in the open ecosystem, a renewed rise in greater economic concentration, and a shift to larger but more quantized, multimodal reasoning models.
China's share of downloads surged to 17.1\% in the recent year, surpassing the collective of American model developers for the first time. 
These advances are driven primarily by DeepSeek and Qwen's reasoning models, with their developers capturing 14\% of recent downloads alone.
At the same time, \Cref{fig:temporal_concentration} shows that economic concentration metrics have suddenly started to rise in 2025, as individual Chinese models and developers consolidate greater download numbers.

Model complexity has also increased dramatically.
\Cref{tab:temporal-shifts} shows mean downloaded model size grew to 20.8B parameters (17$\times$ increase), mainly due to advances in quantization and wider adoption of mixture-of-expert architectures (where a subset of parameters are active at a time).
Similarly, adoption of multimodal generation and video generation models, such as WanAI's Wan2.1 series, each rapidly expanded by 3.4$\times$. 
However, this sophistication came with reduced transparency---only 39\% of model downloads are for systems that disclose available training data (down from 79.3\%).
A new infrastructure layer has emerged, with organizations such as lmstudio-community, comfy, and mlx-community (together accounting for over 22\% of recent downloads) focused on re-packing model formats for more efficient training and inference (e.g. quantization) becoming critical intermediaries. 
The USA's share fell to 15.8\% in the recent period, while the ``International/Online'' category reached 23.8\%, suggesting both geographic diversification and the continued rise of online power users as significant ecosystem participants.

\begin{figure*}
    \centering
    \includegraphics[width=\textwidth]{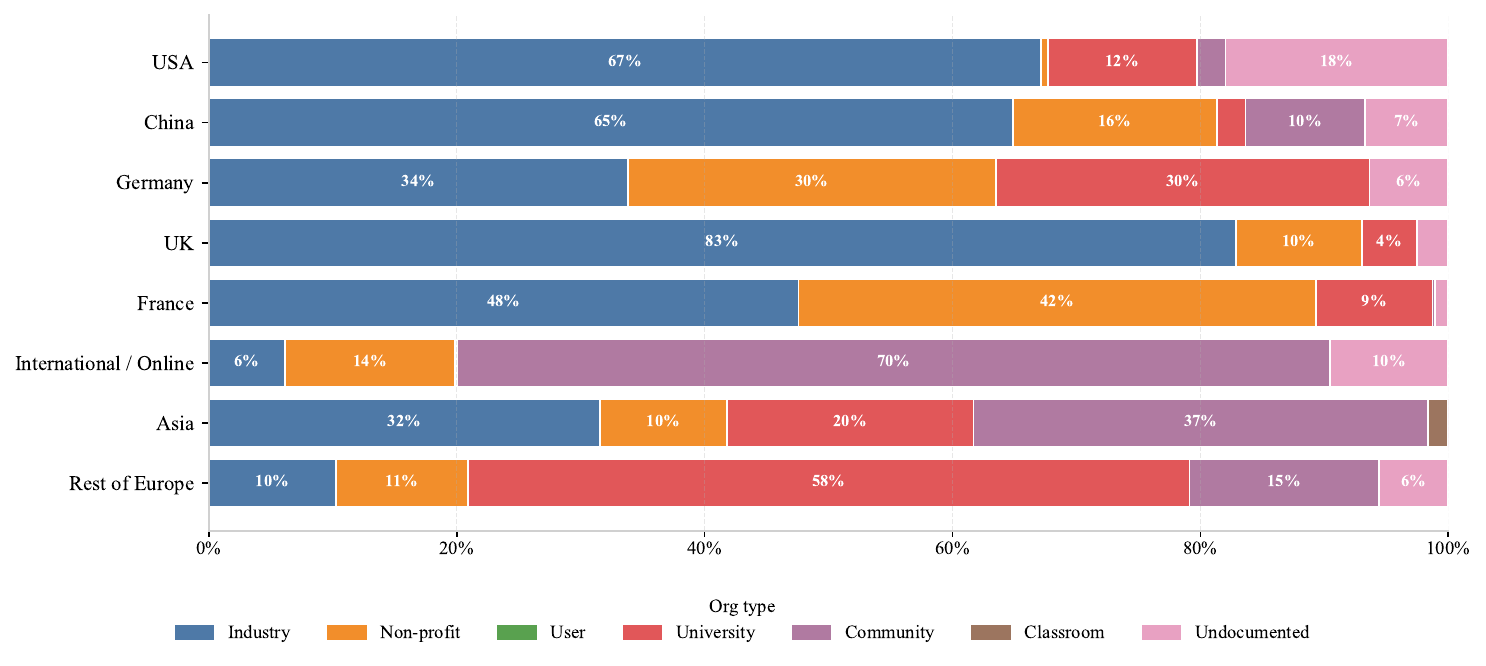}
    \caption{
    The proportion of downloads allocated by developer organization type in each country.
    \textbf{We find US, China, and UK development is skewed heavily to industry,} whereas Germany, France, and the rest of Asia, Europe, and Online development is more balanced towards non-profits, universities, and community contributors.
    }
    \label{fig:org_breakdown}
    \vspace{-2mm}
\end{figure*}

\vspace{-2mm}
\subsection{The Rise of Communities \& Unaffiliated Developers, over Industry Developers}
\vspace{-2mm}

\textbf{Model development and curation is shifting gradually away from large corporations and towards unaffiliated users and online community developers.}
Perhaps most surprisingly, while industry used to occupy 70\% of development before 2022, they now occupy only 37\%.
Most of this has decline has been replaced by independent or unaffiliated users, having risen from 17\% to 39\% in 2025.
However, the most significant upwards trajectory is from what Hugging Face classifies as \emph{Community} developers, that may simply represent international/online volunteer teams, and often operate open source repositories.
Since prior to 2025, their share of downloads has risen 1.5x.
University and non-profit contributors have remained stable in their share over the last few years.

In \Cref{fig:org_breakdown} we breakdown the share of downloads allocated to organizations of different types from each country.
This analysis enables us to understand what types of organizations control open model production by country.
It is immediately clear that whereas US, Chinese, and UK downloads are dominated by industry, other European and Asian countries, as well as Online contributors, have a healthier distribution of non-profits, universities, and community contributors.



\vspace{-2mm}
\section{Shifts in Open Model Characteristics}
\label{sec:shifts}
\vspace{-2mm}

The last five years of open models has seen nuanced shifts in the popularity of model architectures, modalities, training methods, documentation, and access.
We collected supplementary information on these model attributes, beyond what is available in Hugging Face metadata, for the \headsample{} (see \Cref{app:annotation-taxonomy-details}), and then ranked each model attribute by the largest relative shift from before January $2025$ to after January $2025$.
In \Cref{tab:temporal-shifts} we show the largest positive and negative shifts in the types of Model Developers, Model Access, Model Modality, and other Model Attributes.
We devote this section to discussing key shifts in more detail.

\begin{table*}
\centering
\small
\setlength{\tabcolsep}{3pt}
\begin{adjustbox}{width=0.8\textwidth}
\begin{tabular}{c|l|rrrr|r}
\toprule
 & \textsc{Attribute: Category} & $\le$2022 & \textsc{2023} & \textsc{2024} & \textsc{$\ge$2025} & \textsc{$\Delta$\% ($<$2025 | $\ge$2025)} \\
\midrule

\multirow{5}{*}{\rotatebox{90}{\textbf{\textsc{\emph{Organizations}}}}}

& Organization: Community & 0.0\% & 7.9\% & 3.8\% & 9.6\% & \cellcolor{greencell!67} 1.7× \\
& Organization: User & 16.5\% & 34.0\% & 39.6\% & 38.7\% & \cellcolor{greencell!1} 1.1× \\
\cmidrule{2-7}
& Organization: Non-profit & 4.7\% & 4.1\% & 6.4\% & 5.4\% & \cellcolor{redcell!4} 1.0× \\
& Organization: University & 9.6\% & 9.1\% & 8.5\% & 9.2\% & \cellcolor{redcell!15} 1.0× \\
& Organization: Company & 69.2\% & 44.9\% & 41.7\% & 37.0\% & \cellcolor{redcell!80} 0.8× \\

\midrule
\multirow{10}{*}{\rotatebox{90}{\textbf{\textsc{\emph{Model Access}}}}}

& Data Availability: Not Disclosed & 9.8\% & 23.4\% & 31.6\% & 43.1\% & \cellcolor{greencell!80} 1.7× \\
& Model Gating: Use Conditions + Share Info & 0.0\% & 0.5\% & 1.1\% & 1.2\% & \cellcolor{greencell!76} 1.7× \\
& Data Availability: Disclosed \& Unavailable & 2.1\% & 1.2\% & 2.2\% & 2.9\% & \cellcolor{greencell!67} 1.6× \\
& License: Attribution & 0.2\% & 0.6\% & 3.3\% & 3.0\% & \cellcolor{greencell!64} 1.6× \\
& Model Gating: Accept Conditions & 0.0\% & 1.0\% & 3.3\% & 2.4\% & \cellcolor{greencell!22} 1.2× \\
& License: Undocumented & 17.3\% & 29.5\% & 27.7\% & 31.2\% & \cellcolor{greencell!17} 1.1× \\
\cmidrule{2-7}
& License: Open Use & 81.4\% & 54.1\% & 58.3\% & 55.1\% & \cellcolor{redcell!17} 0.9× \\
& License: Acceptable Usage Policy & 0.0\% & 14.5\% & 9.7\% & 9.7\% & \cellcolor{redcell!20} 0.9× \\
& Data Availability: Disclosed \& Available & 79.3\% & 58.5\% & 53.5\% & 39.8\% & \cellcolor{redcell!80} 0.7× \\

\midrule
\multirow{10}{*}{\rotatebox{90}{\textbf{\textsc{\emph{Model Modality}}}}}

& Multimodal Generation & 0.0\% & 0.8\% & 1.3\% & 3.3\% & \cellcolor{greencell!80} 3.4× \\
& Video Generation & 0.0\% & 0.4\% & 0.8\% & 1.8\% & \cellcolor{greencell!79} 3.4× \\
& Text Generation & 21.0\% & 11.0\% & 12.5\% & 20.3\% & \cellcolor{greencell!22} 1.7× \\
& Undocumented & 1.9\% & 7.1\% & 9.7\% & 10.6\% & \cellcolor{greencell!13} 1.4× \\
& Tabular Models & 0.1\% & 0.0\% & 0.1\% & 0.1\% & \cellcolor{greencell!12} 1.4× \\
& Audio Models & 1.0\% & 7.4\% & 9.0\% & 9.2\% & \cellcolor{greencell!6} 1.2× \\
& Multimodal Embedding & 0.3\% & 2.1\% & 2.1\% & 2.1\% & \cellcolor{greencell!3} 1.1× \\
\cmidrule{2-7}
& Image Generation & 0.3\% & 27.9\% & 25.0\% & 23.0\% & \cellcolor{redcell!7} 1.0× \\
& Text Embed/Class & 75.2\% & 28.5\% & 26.9\% & 22.9\% & \cellcolor{redcell!50} 0.7× \\
& Image Embedding & 0.2\% & 14.8\% & 12.5\% & 6.6\% & \cellcolor{redcell!80} 0.5× \\

\midrule
\multirow{16}{*}{\rotatebox{90}{\textbf{\textsc{\emph{Model Attributes}}}}}

& Model Size & 215M & 827M & 1.76B & 20.8B & \cellcolor{greencell!80} 17.0× \\
& Architecture: Mixture-of-Experts & 0.0\% & 0.1\% & 0.4\% & 1.6\% & \cellcolor{greencell!31} 7.2× \\
& Methods: Quantization & 0.0\% & 1.1\% & 3.8\% & 12.2\% & \cellcolor{greencell!22} 5.3× \\
& Methods: RLHF & 0.1\% & 0.5\% & 1.4\% & 4.1\% & \cellcolor{greencell!17} 4.5× \\
& Architecture: Transformer: Text Decoder-only & 8.1\% & 7.4\% & 10.2\% & 20.8\% & \cellcolor{greencell!7} 2.4× \\
& Methods: Instruction finetuning & 0.1\% & 2.5\% & 4.1\% & 6.5\% & \cellcolor{greencell!6} 2.2× \\
& Architecture: Transformer: Speech & 0.8\% & 3.8\% & 6.9\% & 7.7\% & \cellcolor{greencell!2} 1.4× \\
& Methods: Parameter-efficient finetuning & 0.7\% & 6.1\% & 6.5\% & 8.8\% & \cellcolor{greencell!2} 1.4× \\
& Methods: Pretraining: Contrastive Learning & 2.1\% & 5.8\% & 11.7\% & 10.2\% & \cellcolor{greencell!1} 1.2× \\
& Architecture: Diffusion-based Network & 0.1\% & 20.0\% & 18.9\% & 19.5\% & \cellcolor{greencell!1} 1.1× \\
\cmidrule{2-7}
& Architecture: VAE / GAN & 0.0\% & 5.1\% & 2.6\% & 2.8\% & \cellcolor{redcell!24} 0.9× \\
& Architecture: Transformer: Text Encoder-Decoder & 11.8\% & 5.3\% & 5.5\% & 4.8\% & \cellcolor{redcell!40} 0.8× \\
& Architecture: CNN & 0.1\% & 11.3\% & 14.6\% & 8.8\% & \cellcolor{redcell!46} 0.8× \\
& Architecture: Transformer: Text Encoder-only & 76.8\% & 43.4\% & 37.7\% & 32.8\% & \cellcolor{redcell!54} 0.7× \\
& Architecture: LSTM / GRU & 0.4\% & 0.2\% & 0.2\% & 0.2\% & \cellcolor{redcell!60} 0.7× \\
& Methods: Model Merging & 0.0\% & 3.8\% & 1.4\% & 1.3\% & \cellcolor{redcell!73} 0.6× \\

\bottomrule
\end{tabular}
\end{adjustbox}
\vspace{1mm}
\caption{
\textbf{The incidence rate of model attributes over time, weighted by downloads.}
For each time segment, we measure the percent of downloads associated with models that have each \textsc{Attribute: Category} characteristic. 
The $\Delta$\% ($<$2025 | $\ge$2025) column shows the multiplier change in a model attribute from before 2025 to during 2025.
We only include attributes in the table which show a statistically significant shift in this period ($p<0.001$), as measured by a chi-squared test.
}
\label{tab:temporal-shifts}
\vspace{-4mm}
\end{table*}



\vspace{-2mm}
\subsection{The Decline of Open Source}
\vspace{-2mm}

\textbf{Open Source models are on the decline. Model gating and restrictions are on the rise.}
\Cref{tab:temporal-shifts} shows a clear decline in both the availability, and disclosure of a models' training data.
The Open Source Initiative defines open source AI models as those which have open model weights, but also ``sufficiently detailed information about their [training] data.''\footnote{\url{https://opensource.org/ai/open-source-ai-definition}}
Without training data disclosure, a released model is considered ``open weight'' rather than ``open source.''
Whereas in 2022 over 79\% of downloads were for models which disclosed their training data, in 2025 that value is only 39\% (see \Cref{fig:architecture_methods_openness} for more details).

Similarly, model access has become more restrictive, both in terms of gating models (requiring users to accept conditions or share their information first), and more restrictive, tailored, licensing.
Many major models, such as Meta's Llama series, are now gated, comprising over 3.6\% of all model downloads.
As for model licenses, a smaller portion of downloads are allocated to models that provide any license information, as compared to prior years.
And when licenses are documented, Open Use licenses are on the steep decline, whereas non-commercial, or attribution requirements are on the rise (see \Cref{fig:licenses} for more details).

\begin{figure*}
    \centering
    \includegraphics[width=0.8\textwidth]{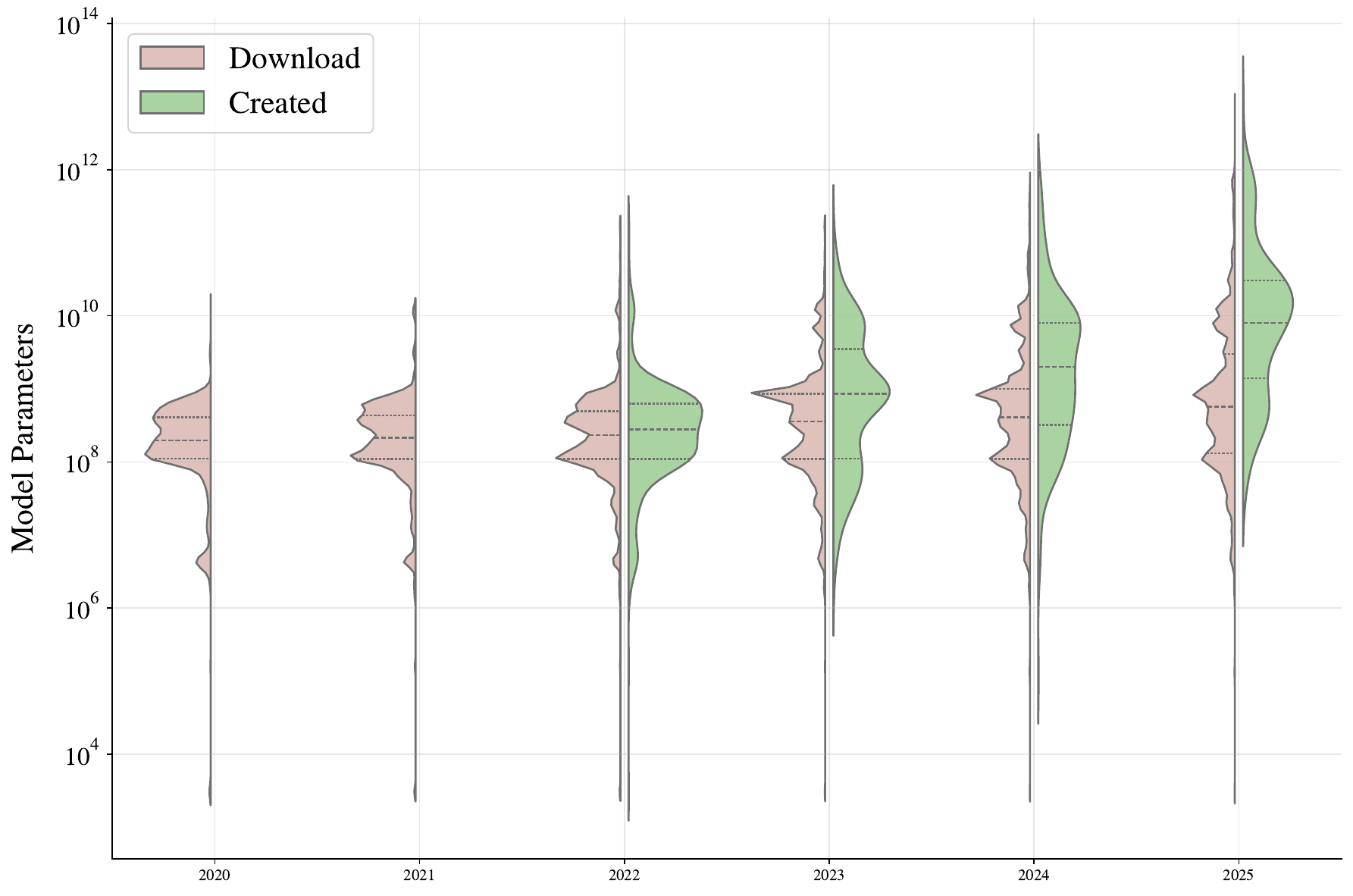}
    \caption{
    \textbf{The distribution of model sizes downloaded in each year (left---pink), and created in each year (right---green) is shifting over time.}
    The created statistics only begin in 2022, as Hugging Face did not record prior model creation times.
    We log-scale the downloads distribution prior to the violin plots' kernel density estimation to smooth and improve the visibility of the various model size peaks.
    The lines within each distribution represent the 25, 50, and 75 percentiles.
    We find the mean size of model download and creation are both rising, though the medians far less quickly as seen in \Cref{tab:temporal-shifts}.
    }
    \label{fig:model_size_static}
    \vspace{-4mm}
\end{figure*}

\vspace{-2mm}
\subsection{Rising Model Sizes}
\vspace{-2mm}

\textbf{The average size of a downloaded model is increasing rapidly, with rising compute availability, advances in quantization, and mixture-of-experts architectures.}
In \Cref{fig:model_size_static} we see the distribution of model sizes according to their download and creation prevalence over time.
First, the mean size of a downloaded model has increased from 827M in 2023 to 20B in 2025.
Note that the median size of a downloaded model remains much lower: 326M in 2023, and 406M in 2025.
This suggests that most of the community's compute affordances have risen more modestly, whereas some power users have scaled to multi-billion parameter architectures.
Much of this power scaling appears to be enabled by a few innovations: aggressive quantization for inference-time memory savings, and mixture-of-expert architectures that have massive parameter counts but many fewer that are active at a time.
For instance, Kimi K2 \citep{kimiteam2025kimik2openagentic} is a popular 1T parameter mixture-of-expert model, with only 32B active per token.
Other popular hundred-billion parameter mixture-of-experts examples include the DeepSeek-V3, Mixtral, Grok-1, MiniMax, and Snowflake Arctic series.
Comparably, Nvidia's Nemotron series comprise fully dense networks in the hundreds of billions of parameters.
The growing adoption to these massive models demonstrates a continued reliance on scaling to achieve performance.

\textbf{The mean size of created models consistently outpaces the mean size of downloaded models, suggesting developers are investing in larger models before most deployers have inference capacity to use them.}
\Cref{fig:model_size_static} shows not only that the mean size of created models is larger than the mean size of downloaded models, but that it is rising much more quickly.
This clearly suggests developers are scaling faster than most deployers can reasonably adopt.
The time lag between development and deployment may span over a year.

\begin{figure*}
    \vspace{-2mm}
    \centering
    \includegraphics[width=\textwidth]{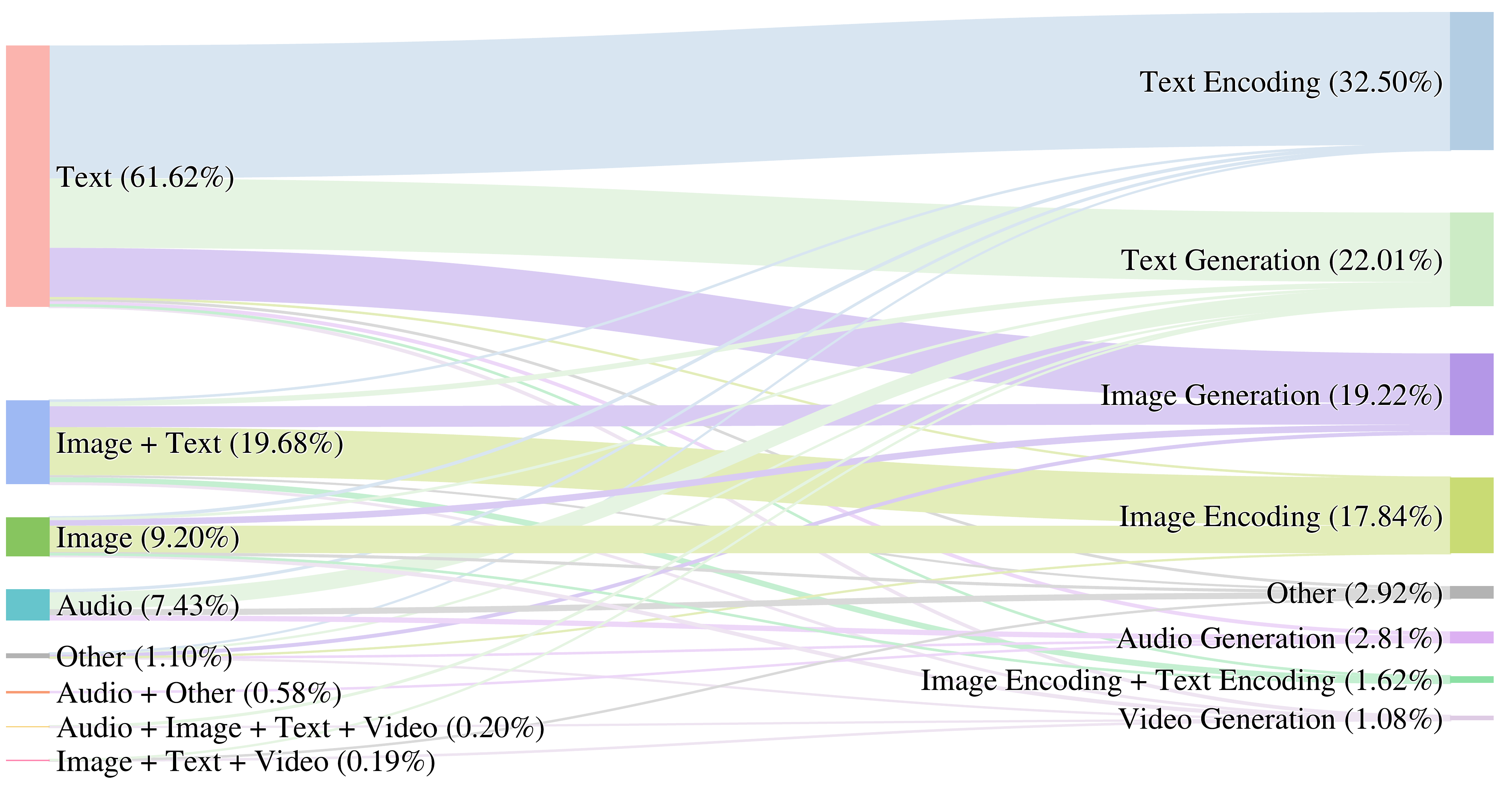}
    \caption{We illustrate the \headsample{}'s aggregate model modality distributions, weighted by allocation of downloads. The model input modalities are depicted on the left, and the model output modalities on the right.
    We find model inputs are predominantly text (61\%), followed by text and images together (20\%), then images (9\%), audio (7\%), followed by other combinations.
    The outputs are ranked as text encoding (33\%), text generation (22\%), image generation (19\%), image encodings (18\%), audio generation (3\%), and video generation (1\%).
    \textbf{We find the output modalities are more heterogeneously distributed than that input modalities on aggregate.}
    }
\label{fig:modalities}
\vspace{-4mm}
\end{figure*}

\vspace{-2mm}
\subsection{The Shift to Generative Multimodal Architectures}
\vspace{-2mm}

\textbf{The ecosystem is quickly migrating towards multimodal, generative systems, and away from encoder-based discriminative architectures.}
\Cref{tab:temporal-shifts} shows two clear temporal trends in the model modality and architecture shifts. 
First, we see a shift in model architectures. 
Encoder-based models, typically used to produce embeddings for retrieval or classification tasks are declining in favor of decoder-based generative models.
Similarly, convolutional neural networks (CNNs), long-short-term-memory cells (LSTMs), and generative-adversarial networks (GANs) are falling off in popularity.
Second, there is a striking rise in the adoption of mixture-of-experts architectures, and for a greater variety of multimodal outputs, such as 3.4x for video).
Increasingly, models incorporate combinations of text, images, audio, and video generation in their output capabilities.
This emphasis on output capabilities even over input capabilities is illustrated in \Cref{fig:modalities}, where output modality heterogeneity is higher than input modalities.

\textbf{Training and inference efficiency methods are becoming an essential stage of model development.}
As deployers scale to larger models, finetuning-adaptation and inference efficiency become paramount to support these systems.
Unsurprisingly, the incidence of parameter-efficient finetuning (PeFT) \citep{pmlr-v97-houlsby19a, pfeiffer-etal-2020-adapterhub} and quantization \citep{nagel2021whitepaperneuralnetwork, MLSYS2024_42a452cb} have increased by 40\% and 500\% respectively since prior to 2025. 
While model merging \citep{yang2024modelmergingllmsmllms} and knowledge distillation \citep{hinton2015distilling, liang2021mixkd} methods have not risen in popularity, reinforcement learning and instruction tuning (RLHF) \citep{NEURIPS2022_b1efde53}, which often leverage larger ``teacher'' models, have become significantly more common-place in terms of usage since 2024.
This suggests a shift from certain techniques leveraging teacher models to other techniques leveraging reward models or synthetic data generators.

\section{Related Work}
\label{sec:related-work}

\textbf{Societal and Political Analyses of AI.}
The discourse surrounding the AI ecosystem is modulated by dichotomies that obscure deeper power dynamics. Much work has been done in the field of critical algorithm studies to understand the societal and political currents steering AI development and deployment. Particularly relevant to this work are; \citet{lehdonvirta:2024:north-south}'s documentation of a "Compute North vs. Compute South" divide, where computational resources concentrate in wealthy nations and reinforce extractive infrastructures. \citet{crawford:2021:atlas} and \citet{noble2018algorithms}  identify how social hierarchies embedded in technical systems operate under the guise of neutrality while perpetuating existing power structures. The dichotomy between closed AI and open AI reveals similar contradictions: despite open models costing 6x less to deploy.

\textbf{Economic and Concentration Analyses of AI.}
Other work on documenting open model characteristics find that closed models dominate 80\% of usage and 95\% of revenue, exposing how "openness" fails to translate into meaningful access \citep{nagle2025latent}. These divides intersect with institutional shifts documented by \citep{ahmed:2023:growing-influence}, who show 96\% of recent AI advances originate from industry, though \citet{vipra2023computational} argue the fundamental barrier remains computational access rather than institutional affiliation. Together, these overlapping binaries fragment understanding while power concentrates across multiple dimensions simultaneously. These divisions mask how advantage compounds through control of compute, data, and talent markets. Building on these dichotomies, the open/closed binary might suggest that open model ecosystems represent a democratic alternative to concentrated power in closed systems, yet empirical studies reveal a more complex reality. 

\textbf{Geopolitical Analyses of AI.}
Work documenting these ecosystems includes Lambert's analyses of China's open-source trajectory and model adoption patterns \citep{Lambert2025ChinaAI,Lambert2025ATOM}, a NIST evaluation of the Chinese DeepSeek models \citep{CAISI2025_DeepSeekEvaluation}, and Epoch AI's study of open models \citep{epoch2024openmodelsreport}. These empirical platform studies demonstrate that openness alone does not distribute power: computational resources, data access, and technical expertise create hierarchies within open ecosystems --- highlighting the importance of studying said ecosystems.

\textbf{Hugging Face Ecosystem Analyses.}
Finally, a growing body of research uses Hugging Face as a testbed for studying open model ecosystems at scale. Ecosystem analyses of the Hugging Face Hub examine the landscape of over 2 million models \citep{laufer2025anatomymachinelearningecosystem, horwitz2025chartatlasworldsmodels} and analyze community dynamics \citep{ait:2023:hfcommunity-tool-analyze-huggingface-hub-community}. Research on model evolution tracks maintenance patterns and versioning challenges \citep{castano:2024:how-machine-learning-models-change,ajibode:2025:towards-semantic-versioning-open-pretrained-language-model-releases}, while studies of reuse practices reveal naming conventions, defects, and ecosystem dependencies \citep{jiang:2023:empirical-study-pretrained-model-reuse-huggingface-registry,Jiang_2025,yang2024ecosystemlargelanguagemodels}. At large, systematic reviews synthesize existing Hugging Face research and validate the platform's suitability as an empirical research site \citep{jones:2024-what-we-know-hf, ait:2025:suitability-huggingface-hub-empirical-studies}. These studies collectively demonstrate Hugging Face's emergence as critical infrastructure both for hosting the open model ecosystem and for understanding its development dynamics.

\section{Discussion} 

Our longitudinal analysis of 2.2B model downloads across 851k models exposes a rapidly evolving open AI ecosystem; one that is simultaneously decentralizing and consolidating, expanding in technical capacity while narrowing in transparency. These trends reveal important structural tensions in how open AI develops, who participates, and which actors come to exercise influence over global model distribution.

\paragraph{Rebalancing of Power and the Cycles of Decentralization.}
Between 2021 and 2024, the open model economy underwent a marked diffusion of power. The download share of the three dominant U.S. industry such as Google, Meta, and OpenAI declined sharply, falling from peaks of >40–60\% to a marginal position by 2025. Unaffiliated developers and loosely coordinated online communities became the primary drivers of model development, in some periods accounting for more than half of all downloads. This shift coincided with the explosive adoption of diffusion-based generative models and LoRA-based customization workflows, which dramatically lowered the barrier to participation.

However, 2025 shows signs of recentralization, driven not by U.S. incumbents but by rising Chinese developers. DeepSeek and Qwen together captured 14\% of all downloads in the most recent year, and China’s national share rose to 17.1\%, surpassing the United States (15.7\%) for the first time.
This pattern of diffusion following a technological shock, then consolidation around leaders of the next technological wave suggests a cyclical rather than linear evolution of the ecosystem.

\paragraph{The Rise of Intermediary Developers.}
A defining structural transformation is the emergence of a new intermediary layer: organizations specializing not in training base models, but in re-packaging, quantizing, adapting, and refactoring them for community use. Groups such as lmstudio-community, comfy, and mlx-community now represent over 22\% of downloads in the most recent year. These intermediaries play an infrastructural role analogous to cloud providers in traditional computing: they translate cutting-edge frontier models into practically deployable artifacts.

This development reflects the industrialization of open model reuse. As model sizes increase from on average from 827M parameters in 2023 to 20B in 2025, efficient inference and adaptation become essential. Intermediaries specialize in this efficiency layer, shaping not only which models become accessible to typical users, but also how innovations diffuse across the ecosystem.

\paragraph{Technical Transformation and Shifting Norms.}
The ecosystem’s technical profile has shifted toward larger, more multimodal, and more computationally efficient architectures. The incidence of video generation and multimodal generation models grew 3.4$\times$, while mixture-of-experts architectures increased 7.2$\times$. Meanwhile, quantization techniques surged 5$\times$, and parameter-efficient finetuning rose by over 40\%, indicating their centrality to contemporary deployment practice.

Importantly, the rise in scale is concentrated in the long-tail of high-capacity deployers: while the mean downloaded model grew to 20B parameters, the median rose only modestly from 326M (2023) to 406M (2025). This divergence indicates expanding inequality between power users capable of hosting multi-billion-parameter models and typical developers who remain limited by compute constraints.

\paragraph{Declining Transparency.}
Despite rapid growth in model availability, openness has deteriorated substantially. Models disclosing any information about training data fell from 79.3\% (2022) to 39\% (2025), and for the first time, open-weights models outnumber truly open-source models. At the same time, model gating, attribution requirements, and use-condition licenses all increased.

This shift reflects real tensions: as models grow larger and more commercially valuable, incentives to restrict access intensify. Yet the decline in transparency raises profound questions for reproducibility, governance, and accountability; precisely in the period when open models are absorbing heightened geopolitical and economic relevance.

The rapidity of these structural shifts underscores the need for continuous monitoring of the open model ecosystem. Traditional static snapshots of the AI landscape fail to capture the month-to-month dynamics of consolidation and participation. The combination of rising national competition, opaque licensing practices, and fast-paced architectural innovations demands tools analogous to competition-authority monitoring in other economic sectors. Our dataset and dashboard provide one such mechanism, enabling researchers and policymakers to track concentration, transparency, and participation longitudinally.

\section*{Ethics Statement}
\label{sec:ethics}
\textbf{Limitations.}
It is important to note that Hugging Face download metrics, or any other metrics, are at best proxies for real usage. 
We make the case in \Cref{sec:methodology} for why we believe this data, with sufficient de-duplication and filters, provides the most accurate estimate of model usage, though it still remains imperfect.
As a result, these metrics best serve the analysis of broader trends and relative popularity over time, rather than precise numerical designations of economic concentration.

\textbf{Privacy.}
This study analyzes aggregate download and metadata statistics from the publicly available Hugging Face Model Hub to understand structural shifts in the open AI ecosystem. All data used are non-personal and pertain exclusively to publicly released AI models, their metadata, and anonymized download counts. No individual user identifiers or personal information were accessed or retained.
We decided to only identify organizations' countries, not for individual users, to retain anonymity.

\textbf{Artifact Release.}
By releasing the public dashboard and annotation methodology openly, we aim to enable reproducibility, critical analysis, and further community research into market concentration, openness, and data transparency. The data and code are shared under an open license consistent with responsible research and privacy guidelines.

\textbf{Responsible and fairly compensated data labor.}
We recognize the importance of fair compensation and protections for data laborers in AI~\cite{du2025reimagining}. Annotators for this project were employed through Upwork and received \$25/hour--well above average U.S. minimum wage~\cite{usdol2025minimum}. Strict timelines were not imposed and annotators were able to determine and work according to their own schedules. Additionally, we value the technical expertise and diverse perspectives our annotators contributed and continuously improved and expanded our taxonomy with their input. We believe transparency regarding data labor practices is essential and all annotators have accepted the offer to be recognized for their work in the acknowledgments.

\textbf{Framing.}
Finally, we acknowledge potential ethical risks associated with framing AI development through geopolitical or corporate competition. Our analysis seeks to document ecosystem dynamics, not to reinforce nationalistic narratives or proprietary advantage. We encourage subsequent work to combine quantitative analysis with qualitative inquiry into the social and cultural dimensions of open AI ecosystems.

\section*{Acknowledgments}

We would like to thank Emily Wenger, Nathan Lambert and Florian Brand from the ATOM Project, and Jaime Sevilla from Epoch AI, for their guidance and perspectives. We would also like to thank our model annotators for their efforts and diligence:  Namrah Fahad Karim, Mostafa Ahmed, Noman Jalal, and Goran Vukadinovic. Further, we express our gratitude to Lysandre Debut for the support in collecting the Hugging Face download data.

\clearpage

\bibliographystyle{unsrtnat}
\bibliography{references}

\clearpage


\clearpage
\appendix
\addcontentsline{toc}{section}{Appendix}
\part{Appendix} 
\parttoc
\newpage

\section{Contributions}

\begin{itemize}
    \item \textbf{Shayne Longpre}: Co-led and designed the project with Lucie. Led analysis on economic concentration and on shifting model attributes.

    \item \textbf{Christopher Akiki}: Led the data pipeline, and the model metadata extraction from the Hugging Face platform.
    
    \item \textbf{Campbell Lund}: Co-led the annotation taxonomy design and led the annotation collection process.
    
    \item \textbf{Atharva Kulkarni}: Co-led the data cleaning pipeline. Led the analysis on the temporal distribution of model sizes downloaded vs created, model modality distribution, and model training methods.

    \item \textbf{Emily Chen}: Led the design and implementation of the open model visualization dashboard. Created the global map and leaderboard figures, and implemented an efficient version of the data pipeline.

    \item \textbf{Irene Solaiman}: Project senior advisor.
    \item \textbf{Avijit Ghosh}: Project senior advisor and dashboard implementation advisor.
    \item \textbf{Yacine Jernite}: Project senior advisor.
    
    \item \textbf{Lucie-Aim\'{e}e Kaffee}: Co-led to the design and narrative of the project with Shayne as senior advisor. Led the data collection of the Hugging Face model downloads dataset. Contributed to the Concentration of Power section. 
    
\end{itemize}

\section{Data Sources}
\label{app:dataset-details}

\begin{table}
\small
\setlength{\tabcolsep}{3pt}
\begin{adjustbox}{width=\textwidth}
\begin{tabular}{l|ll|p{14cm}}
\toprule
\textsc{Data} & \textsc{Src} & \textsc{Cov} & \textsc{Description} \\
\midrule
\textbf{Usage Logs} & \hf & All & Weekly downloads for public models from June 2020 to July 2025.  \\
\textbf{Model Size} & \hf \twemoji{robot}  & All & Estimated model parameters, using both safetensors and regressing model file byte size against the parameter counts. \\
\textbf{Model Developer} & \hf \twemoji{pen} & All & The model developer, and their organization's headquarters (including international and online). \\
\textbf{Architecture \& Modalities} & \twemoji{pen} & Head & The model's architecture, and its input and output modalities. \\
\textbf{Training Methods} & \twemoji{pen} & Head & The methods used to train this model, or derive it from another model \\
\textbf{Language(s)} & \hf \twemoji{pen} & All & The languages used to train this model, if text-based. \\
\textbf{Model \& Data Access} & \twemoji{pen} & Head & If the model is gated, and if the training datasets are documented and accessible. \\
\textbf{Model Graph} & \hf \twemoji{pen} & Head & A graph showing what model(s) each model was derived from. \\
\bottomrule
\end{tabular}
\end{adjustbox}
\vspace{1mm}
\caption{\textbf{A list of the data collected, their sources, and coverage.}
We list the source(s), whether from Hugging Face APIs (\hf), automatic crawling (\twemoji{robot}), or manual collection (\twemoji{pen}), and whether the metadata covers \emph{All} models, or just the \emph{Head} sample.}
\label{tab:data}
\end{table}

We summarize our data collection in \Cref{tab:data}.
These data sources are collected from a mix of internal Hugging Face download logs, public model metadata on the Hugging Face hub, as well as manually or automatically inferred model metadata.

\subsection{Hugging Face Model Download Data}
The data on model downloads is publicly available on Hugging Face for the past 30 days for each model. To access download data for all models across all time, we leverage internal databases to collect information on download statistics, aggregated on the weekly level. We filter data to only cover models that are \textit{currently} publicly available, i.e., at the time of data collection, using the Hugging Face API \texttt{list\_models} function.

\subsection{Hugging Face Hub Model Metadata}

The metadata fields that augment each row of the dataset have multiple sources, though most are returned by specific API endpoints of the Hugging Face Hub. The organization metadata is scraped from the Hugging Face website itself since detailed information is not yet available for structured querying. The model cards data column is extracted by a dataset maintained and updated daily by Hugging Face ML librarian Daniel van Strien (https://huggingface.co/datasets/librarian-bots/model\_cards\_with\_metadata). The modalities are automatically extracted using heuristics based on structured data extracted from the Hugging Face documentation and tasks page (https://huggingface.co/tasks). A more detailed and extensive overview is included in the dataset cards of each of the released datasets.

\subsection{Geographical Metadata for Model Developers}

In our annotations for national association, we only focus on organizations, not user accounts, to preserve privacy of individual contributors. 
For each model developer, the authors manually annotated the country associated with the organization. 
This is done by finding evidence of the organization's website on the Hugging Face page.
The country is chosen as the organizations' headquarters.
For certain organizations that are primarily based online, or exist without a primary headquarters based in one nation, we label them as ``Online'', or ``International'', respectively.
For individual contributors to the platform, where no organization appears directly affiliated with the release of the model, we label the origin as ``Individual''.

\section{Detailed Results}
\label{app:detailed-results}

\begin{figure*}[t]             
  \centering
  
  \begin{subfigure}{\textwidth}
    \includegraphics[width=\textwidth,
                     clip,trim=0 10 0 0]{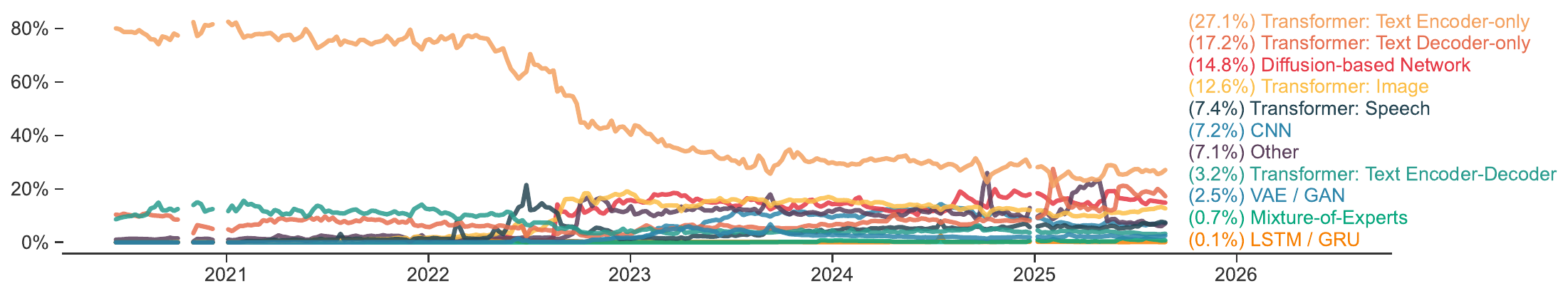}
  \end{subfigure}
  
  
  \begin{subfigure}{\textwidth}
    \includegraphics[width=\textwidth,
                     clip,trim=0 10 0 0]{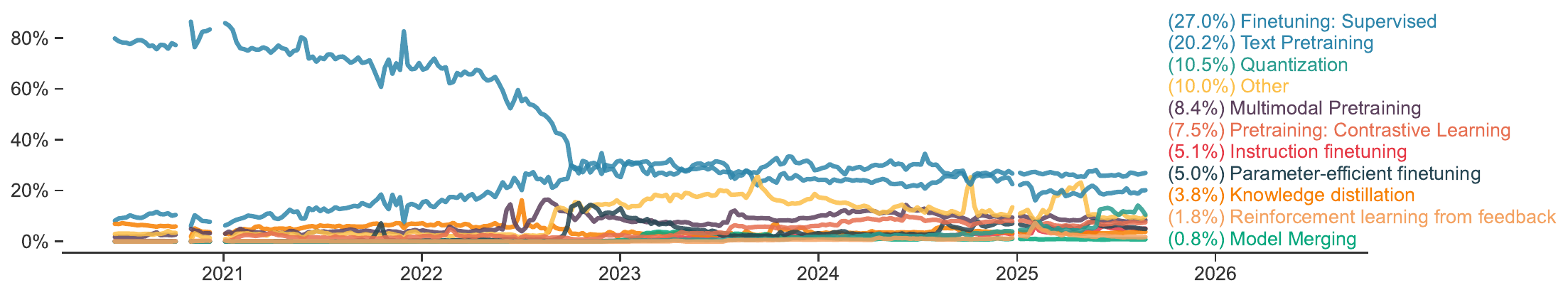}
  \end{subfigure}
  
  
  \begin{subfigure}{\textwidth}
    \includegraphics[width=\textwidth,
                     clip,trim=0 10 0 0]{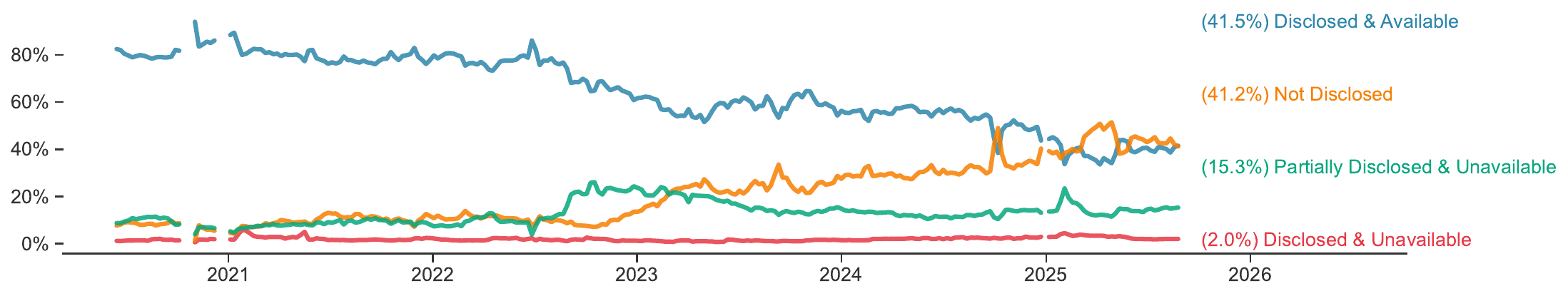}
  \end{subfigure}
  
  \caption{
  These plots measure the portion of downloads associated with different model attributes.
    \textbf{Top:} Model architectures over time.  
    \textbf{Middle:} Training and inference methods over time.  
    \textbf{Bottom:} The disclosure and availability of a models' training data.
    }
  \label{fig:architecture_methods_openness}
\end{figure*}

\begin{figure*}
    \centering
    \includegraphics[width=\textwidth]{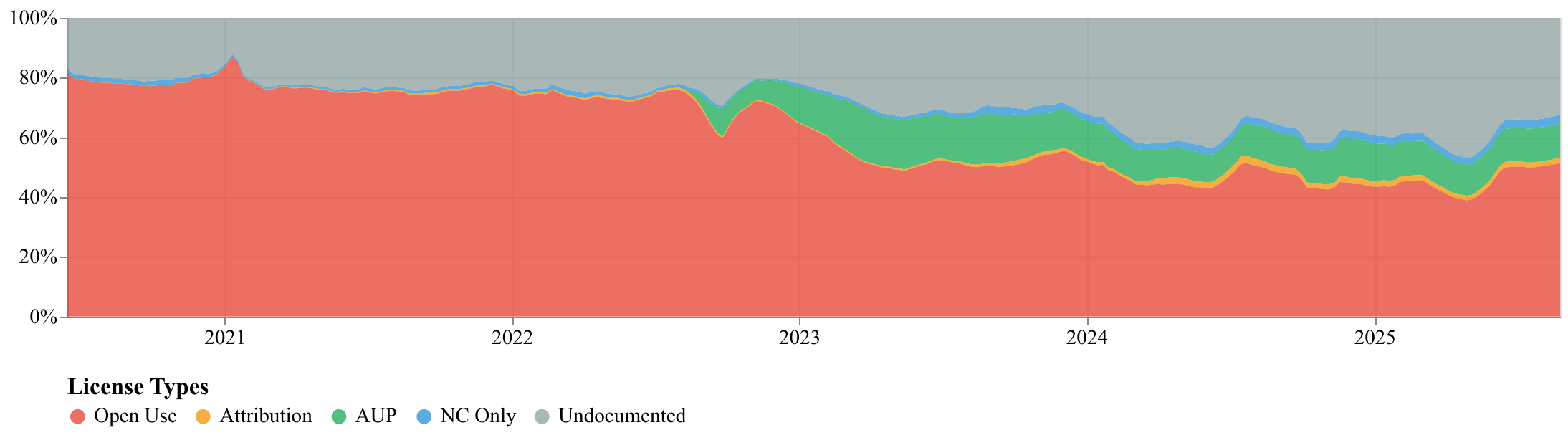}
    \caption{The portion of downloads associated with models of a given license type. \textbf{Open Use licenses are on the decline, mainly replaced by licenses with Acceptable Usage Policies (AUPs).}}
    \label{fig:licenses}
\end{figure*}

\begin{figure*}
    \centering
    \includegraphics[width=\textwidth]{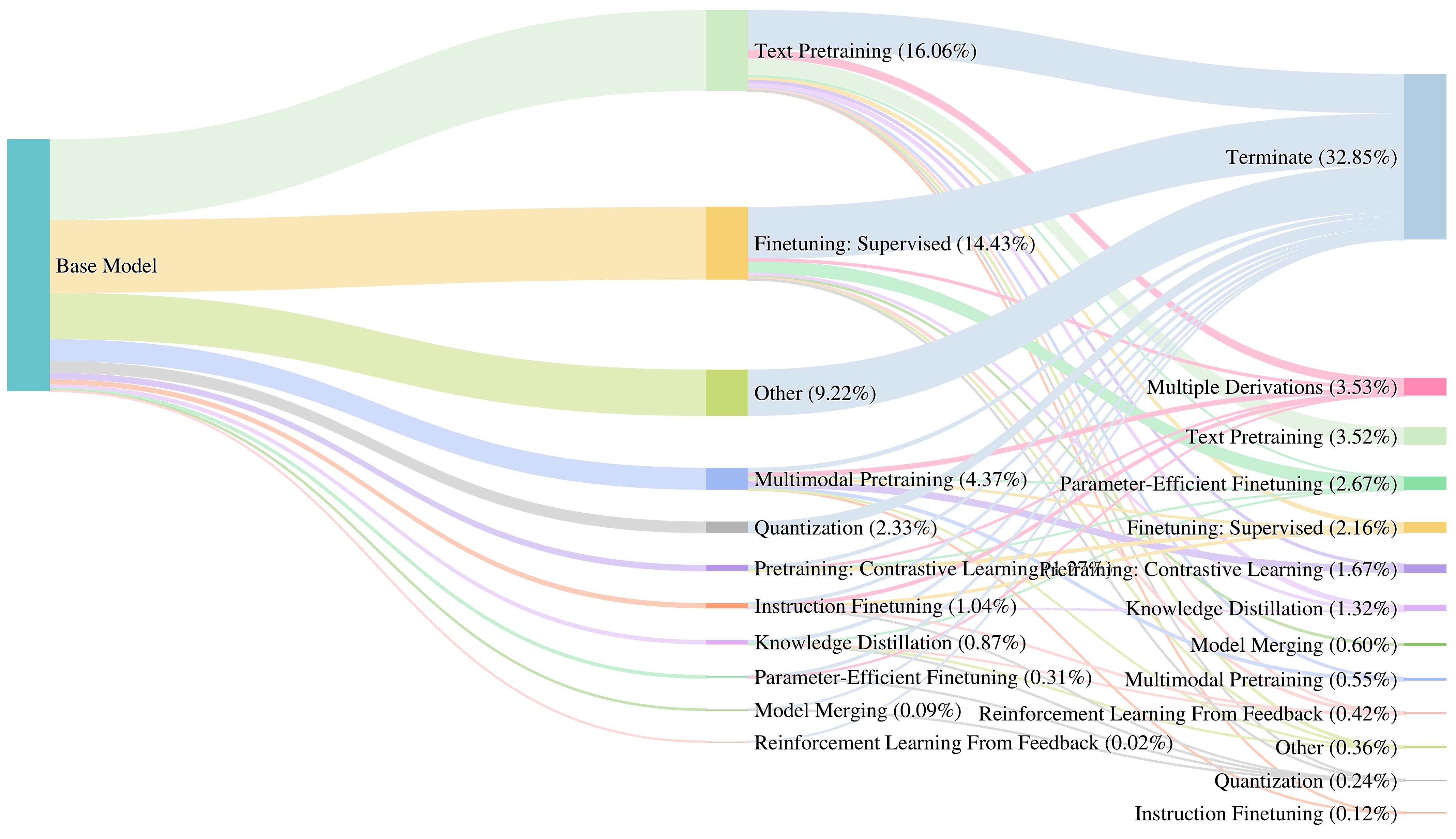}
    \caption{
    We illustrate a network derivation graph, showing the portion of downloads allocated to models derived according to different methods.
    Each column sums to 100\%.
    We find that supervised finetuning is the most popular type of derivation (21\%), followed by continued pretraining (16\%), and quantization (4\%).
    Models that undergoe a second round of derivations see additional pretraining (7\%), parameter-efficient finetuning (5\%), and finetuning (4\%).
    And a full 7\% of model downloads are allocated to systems with 2+ series of derivations.
    }
    \label{fig:training_methods}
\end{figure*}

In \Cref{fig:architecture_methods_openness} we show fine-grained temporal trends for model architecture (top), model development methods (middle), and data disclosure/transparency (bottom).
Additionally, \Cref{fig:licenses} shows how the proportion of license types has changed over time.
Lastly, \Cref{fig:training_methods} shows the download-weighted graph of models that have been derived and in what ways they have been derived.
These figures provide greater details into the temporal trends and model derivations on the Hugging Face hub, as weighted by usage metrics.

\section{Annotation Taxonomy \& Design}
\label{app:annotation-taxonomy-details}

To gain meaningful insight into the specific features and types of models that populate the Hugging Face Ecosystem, we needed to collect information beyond what is available programmatically or documented within model cards. We recruited four annotators to perform this manual analysis: two based in the United States, one in Serbia, and one in Egypt. All annotators had a strong background in computer science/machine learning and participated in training before beginning annotation tasks.

When necessary, we instructed annotators to search for and consult additional sources (e.g., the accompanying publication or associated GitHub repository) in order to answer the following questions about each model:
\begin{itemize}
  \item Is the model gated on Hugging Face?
  \item What are the model input and output modalities?
  \item What languages are the models trained to cover?
  \item Is the model derived from another model?
  \item What is the model architecture?
  \item What training methods were used?
  \item Are the datasets used to train the model fully documented and publicly available?
\end{itemize}

\subsection{Model Gating}
Model authors have the ability to gate access to their models on Hugging Face. Often, users gain access by simply agreeing to share their username and email address with the model author; however, some models require additional information or agreement to specific terms of use. To document which models are open access and which are gated, we have provided the following instructions to our annotators:
\medskip

\begin{promptbox}[Task: Is the model gated on Hugging Face?]
\subsection*{Description}
On Hugging Face, some models can only be accessed or downloaded once some terms have been accepted or information entered. This might be as simple as clicking “Agree and Access Repository” on the main page, or filling out additional information. Please indicate if this is the case.
\medskip

\subsection*{Multi-label Options}
\begin{itemize}
    \item N/A
    \item Accept to share username \& email
    \item Other agreement/info requirements
\end{itemize}
\end{promptbox}

\subsection{Modality}
For each model, we note the associated input and output modalities. For example, generative models may support a single modality (e.g., text → text) or multimodal (e.g., text → image), while other models are trained for specific tasks (e.g., text → classification). This allows us to understand how the priorities of model authors have shifted across modality. We have provided the following instructions to our annotators to document input and output modality:
\medskip

\begin{promptbox}[Task: What is the input and output modality of the model?]
\subsection*{Description}
Identify the data types that the model takes as input and produces as output. There will be two separate columns for this task, one for documenting input modality and one for documenting output.

\begin{itemize}
    \item Models may be a single modality (e.g., text → text) or multimodal (e.g., image → text). 
    \item If the model is not generative, list the output type (e.g., text → classification).
    \item Refer to the model card, and, if available, demo and associated paper.
    \item Use the predefined labels for consistency.
    \item If the model uses a modality not covered by the provided options, select \texttt{Other} and clarify in the Note column.
    \item Write the exact label as formatted in the multi-label options below to ensure the annotations are machine-readable.
    \item If you specify \texttt{Other} use the Note column to explain.
\end{itemize}

\subsection*{Multi-label Input Options}
\begin{itemize}
    \item Text
    \item Image
    \item Video
    \item Speech
    \item Tabular
    \item Other
\end{itemize}

\subsection*{Multi-label Output Options}
\begin{itemize}
    \item Text Generation
    \item Image Generation
    \item Video Generation
    \item Speech Generation
    \item Tabular
    \item Text Classification
    \item Text Sequence Classification
    \item Image Classification
    \item Image Segmentation
    \item Image Bounding Boxes
    \item Text Embedding
    \item Image Embedding
    \item Other
\end{itemize}
\end{promptbox}

\subsection{Languages}
In an effort to measure linguistic inclusion and gain insight into the rate at which language groups are represented across popular Hugging Face models, we document each instance a model is trained or finetuned on a specific language dataset. In addition to natural language, we also note cases where a model is trained on coding languages. We have provided the following instructions to our annotators to document language:
\medskip

\begin{promptbox}[Task: What languages are the model trained to accommodate?]
\subsection*{Description}
Determine which natural and coding languages the model is designed to handle. This typically refers to the language(s) present in the training data for text or speech models. 
\medskip

\begin{itemize}
    \item If language information is unavailable on the model card, check the associated paper, dataset links, or GitHub documentation.
    \item Use ISO language codes (e.g., \texttt{EN} for English, \texttt{DE} for German) where possible.
    \item If multiple languages are supported, document them all in a comma-separated list.
    \item If the model does not generate language but, for example, has pre-defined classification labels, list the language(s) of those labels.
    \item If there are clearly over 50 languages, then simply mark it as \texttt{multilingual}.
    \item If this model was derived from another, then only include the languages used in the derivation process (e.g., fine-tuning), not the base model’s pretraining languages.
\end{itemize}
\end{promptbox}

\subsection{Model Derivation}
A model is considered derived if it inherits the weights of a pre-existing base model before conducting additional training, quantization, or other adaptations. To gain insight into upstream influence and which models power the Hugging Face ecosystem, we have provided the following instructions to our annotators to document derivations:
\medskip

\begin{promptbox}[Task: Is the model derived from another model?]
\subsection*{Description}
Indicate whether the model is derived from another pre-existing model. A model is considered derived if it takes another model’s weights and conducts additional training, quantization, or model merging. A model is not considered derivative if it uses the same architecture or code.
\medskip

\begin{itemize}
    \item This may be mentioned in the model card (often under \texttt{base model}), the repository, or the paper.
    \item Provide the full name of the base model if known.
    \item If the model was trained entirely from scratch, indicate \texttt{No}.
    \item Input the exact Hugging Face model name, so it can be looked up automatically. 
    For example, for BERT it would likely be \texttt{google-bert/bert-base-uncased}.
\end{itemize}
\end{promptbox}

\subsection{Architecture}
The architecture of a model determines how it reasons and processes input. Documenting which architectural choices model authors have implemented across time allows us to identify dominant paradigms (e.g., transformers or diffusion models) and assess how shifts in architecture reflect emerging tasks, modalities, or compute constraints. We have provided the following instructions to our annotators to document model architectures:
\medskip

\begin{promptbox}[Task: What is the model architecture?]
\subsection*{Description}
Specify the architecture used by the model, such as \texttt{Transformer}, \texttt{ResNet}, \texttt{UNet}, \texttt{LSTM}, etc. 
\begin{itemize}
    \item Use precise architecture names as described in the model documentation or paper.
    \item Include whether the model is a variant or combination (e.g., Vision Transformer + Decoder).
    \item If the model encodes multiple modalities, comma-separate the labels. \\
    For example: \texttt{Transformer: Text Encoder-only, Transformer: Image Encoder-only}.
    \item A model might also combine architectures, such as being both a \texttt{Transformer} and a \texttt{Mixture-of-Experts} model.
    \item Common examples have been pre-categorized here.
\end{itemize}
\medskip

\subsection*{Multi-label Options}
\begin{itemize}
    \item Transformer: Text Encoder-only
    \item Transformer: Text Decoder-only
    \item Transformer: Text Encoder-Decoder
    \item Transformer: Image Encoder-only
    \item Transformer: Image Decoder-only
    \item Transformer: Image Encoder-Decoder
    \item Transformer: Speech Encoder-only
    \item Transformer: Speech Decoder-only
    \item Transformer: Speech Encoder-Decoder
    \item Transformer: Unknown
    \item LSTM
    \item GRU
    \item CNN
    \item Diffusion-based Network
    \item Variational Autoencoder
    \item Mixture-of-Experts
    \item Generative Adversarial Network (GAN)
    \item Unsure
    \item Other
\end{itemize}
\end{promptbox}

\subsection{Training Methods}
The training methods used when creating a model impact how the model performs on specific tasks. Documenting which training techniques model authors have opted for over time allows us to trace the adoption of popular new techniques (e.g., fine-tuning, reinforcement learning and quantization), and assess how the field balances the reuse of existing models with the creation of new ones. We have provided the following instructions to our annotators to document model training methods:
\medskip

\begin{promptbox}[Task: What training methods were used?]
\subsection*{Description}
Describe the training strategy applied to the model. This could include pretraining methods (e.g., masked language modeling, contrastive learning), fine-tuning approaches, reinforcement learning (e.g., RLHF), or other techniques (e.g., instruction tuning, distillation).
\medskip
\begin{itemize}
    \item Refer to the model card or original paper.
    \item We propose concise and standard terminology below. If you list \texttt{Other}, then please also suggest what it is.
    \item If this model was derived from another, then only include the training methods used to make the derivation (e.g., fine-tuning, not the base model’s pretraining).
\end{itemize}
\medskip

\subsection*{Multi-label Options}
\begin{itemize}
    \item Pretraining: Masked Language Modeling (MLM)
    \item Pretraining: Causal Language Modeling (CLM)
    \item Pretraining: Denoising Autoencoder
    \item Pretraining: Contrastive Learning
    \item Pretraining: Variational Autoencoder
    \item Pretraining: Multimodal joint-embeddings
    \item Pretraining: Supervised
    \item Pretraining: auto-regressive image generation
    \item Pretraining: Next Sentence Prediction (NSP)
    \item Finetuning: Supervised
    \item Finetuning: Prompt-based tuning
    \item Parameter-efficient finetuning
    \item Knowledge distillation
    \item Instruction finetuning
    \item Multi-task finetuning
    \item Reinforcement learning from feedback
    \item Adversarial Training
    \item Quantization
    \item Model Merging
    \item Unsure
    \item Undisclosed
    \item Other
\end{itemize}
\end{promptbox}

\subsection{Training Data Availability}
An essential component of transparent model development is data provenance. In order to monitor trends in openness, we document whether training datasets are publicly available or proprietary. We have provided the following instructions to our annotators to document training data availability:
\medskip

\begin{promptbox}[Task: Are the datasets used to train the model disclosed and publicly available?]
\subsection*{Description}
Find out if the Hugging Face model card or associated paper for the model disclosed what training datasets were used. We would also like to know if those datasets are publicly available or proprietary. 
\medskip
If this model was derived from another, then we only care about the training methods used to make the derivation (eg. fine-tuning datasets but not the base model’s pre-training datasets).
\medskip
\begin{itemize}
    \item \textbf{Not disclosed} — There is no information on the datasets used at all.
    
    \item \textbf{Partially disclosed: unavailable} — General information is provided (e.g., \texttt{web data}) but nothing else.
    
    \item \textbf{Disclosed: unavailable} — A detailed list of the dataset sources is provided, but they are proprietary.
    
    \item \textbf{Disclosed: available} — The dataset(s) are disclosed, named, and public. \\
    In this case, please add a note on where you found the dataset(s) listed. \\
    All of the main datasets must be publicly available for it to count as \texttt{Disclosed: available}.
    
    \item If the model is derived, only consider the data used for the derivative. \\
    For example, if the model is a derivative of \texttt{google-bert/bert-base-uncased} (which has publicly disclosed and available training data) but does not mention any additional data, label it as \textbf{Not disclosed}.
\end{itemize}
\medskip

\subsection*{Label Options}
\begin{itemize}
    \item Not disclosed
    \item Partially disclosed: unavailable
    \item Disclosed: unavailable
    \item Disclosed: available
\end{itemize}
\end{promptbox}

\end{document}